\documentclass[11pt,letterpaper]{article}
\usepackage{preprint}
\usepackage{multicol}
\usepackage{xurl}
\providecommand{\BIBdoi}[1]{\href{https://doi.org/#1}{\nolinkurl{#1}}}

\title{From Splats to Silicon: A Workload-Centric Survey of Efficient 3D Gaussian Splatting}
\author{Minnan Pei\textsuperscript{1,*}, Qiwei Dong\textsuperscript{2,*},
Yihan Zhou\textsuperscript{1}, Gang Li\textsuperscript{1,$\dagger$},
Yuchen Zhu\textsuperscript{3}, Wenju Zhao\textsuperscript{4},
Zhongtian Long\textsuperscript{4}, Siting Wang\textsuperscript{1},
Zeyu Zhu\textsuperscript{1}, Peisong Wang\textsuperscript{1}, and
Jian Cheng\textsuperscript{1,$\dagger$}}
\date{}
\hypersetup{
  pdftitle={From Splats to Silicon: A Workload-Centric Survey of Efficient 3D Gaussian Splatting},
  pdfauthor={Minnan Pei, Qiwei Dong, Yihan Zhou, Gang Li, Yuchen Zhu, Wenju Zhao, Zhongtian Long, Siting Wang, Zeyu Zhu, Peisong Wang, Jian Cheng},
  pdfsubject={3D Gaussian splatting system efficiency}
}

\begin{document}
\preprintbibcontrol{IEEErefstyle}
\twocolumn[{%
\begin{minipage}{\textwidth}
\maketitle

{\small\raggedright
\textsuperscript{1}Institute of Automation, Chinese Academy of Sciences, and University of Chinese Academy of Sciences\par
\textsuperscript{2}Nanjing University
\hspace{0.75em}\textsuperscript{3}Eindhoven University of Technology\par
\textsuperscript{4}School of Computer Science and Technology, Huazhong University of Science and Technology\par
\smallskip
\textsuperscript{*}Equal contribution.
\enspace\textsuperscript{$\dagger$}Gang Li and Jian Cheng are co-corresponding authors.\par
\href{mailto:peiminnan19@mails.ucas.ac.cn}{\nolinkurl{peiminnan19@mails.ucas.ac.cn}},
\href{mailto:gang.li@ia.ac.cn}{\nolinkurl{gang.li@ia.ac.cn}},
\href{mailto:qiweidong@smail.nju.edu.cn}{\nolinkurl{qiweidong@smail.nju.edu.cn}}\par
}

\begin{abstract}
3D Gaussian splatting (3DGS) represents scenes with explicit Gaussians for interactive view synthesis, yet its efficiency varies widely across scenes, viewpoints, resolutions, and platforms. Existing studies optimize algorithms, GPU runtimes, and architectural support, but isolated metrics such as model size, Gaussian counts, and local speedups rarely predict end-to-end performance. This survey unifies all three layers along a shared rendering and update path through a workload-centric framework tracing five counts from resident Gaussians through active sets, Gaussian--tile associations, and pixel evaluations to blended contributions. Pairing this synthesis with controlled GPU profiling under matched inputs and quality bounds across scenes, resolutions, trajectories, and backends reveals three concrete cross-layer findings. First, fewer associations or smaller compressed models can still increase latency or DRAM traffic when temporary attribute copying or per-frame decoding offsets geometric savings. Second, tiling granularity balances opposing stage costs, while fixed screen-space culling exposes resolution- and trajectory-dependent quality and tail-latency trade-offs. Third, faster rasterization shifts bottlenecks toward updates, memory transfers, and cross-frame recovery, while unaligned rendering defaults can silently confound comparisons. We conclude with evaluation principles and open directions; our profiling toolkit GSPROF is available at \url{https://github.com/HAPPYPMN/gsprof}.
\end{abstract}

\par
\end{minipage}%
}]
\thispagestyle{plain}

\section{Introduction}
\label{sec:intro}
\label{sec:introduction}

3D Gaussian splatting (3DGS) replaces the dense ray sampling and neural network queries of neural radiance fields (NeRFs) with explicit anisotropic 3D Gaussians, enabling interactive novel-view synthesis across extended reality, digital twins, free-viewpoint media, and robotic perception~\cite{ref01,ref02,ref99,ref83,ref84}. By exposing scene geometry and appearance directly, this representation extends 3DGS from static rendering to dynamic scenes, large-scale reconstruction, and online mapping, making efficient execution across heterogeneous platforms a central systems challenge~\cite{ref02,ref81,ref80,ref83}. At the same time, explicit Gaussians shift rendering cost from regular network evaluation to irregular, view-dependent dataflow processing across Projection, Sorting, and Blending. A scene commonly contains hundreds of thousands to millions of Gaussians whose visibility, screen-space footprints, Gaussian-tile associations, and per-pixel blending depth fluctuate sharply with scene scale, camera proximity, and output resolution~\cite{ref02,ref38,ref44}. Training and online updates further add gradient accumulation, optimizer states, and host-device transfers~\cite{ref83,ref73,ref61,ref62}.

To address these costs, efficiency research has expanded across three layers: algorithms compact stored attributes, prune Gaussians, select levels of detail, or modify compositing rules; GPU runtimes refine spatial partitioning, warp scheduling, and multi-tier data placement; and architectural support extends graphics pipelines or introduces specialized units for sorting, buffering, and near-memory evaluation~\cite{ref31,ref38,ref44,ref73}. Across these layers, however, reported metrics reflect different points along the execution path rather than end-to-end system performance. Compression studies evaluate model size without tracking whether decoding inflates resident memory or slows per-frame rendering; GPU runtime studies report fewer Gaussian-tile associations or local kernel speedups even when intermediate attribute gathering offsets net speedup; and hardware studies report area and throughput on isolated stages while leaving other stages or parameter updates on a host processor~\cite{ref38,ref73,ref32,ref40,ref37}. 
\begin{table*}[!t]
\caption{Comparison with Existing 3DGS Surveys}
\label{tab:survey-comparison}
\centering\scriptsize
\setlength{\tabcolsep}{3pt}
\renewcommand{\arraystretch}{1.14}
\newcommand{\primary}{\scalebox{1.2}{$\bullet$}}
\newcommand{\partialscope}{$\triangle$}
\newcommand{\outside}{\scalebox{1.3}{$\circ$}}
\newcommand{\surveyheading}[2][c]{\begin{tabular}[c]{@{}#1@{}}#2\end{tabular}}
\begin{tabularx}{\textwidth}{@{}>{\raggedright\arraybackslash}p{.146\textwidth}Yccccc@{}}
\toprule
\surveyheading[l]{Survey Focus} & \surveyheading[l]{Representative Works} & \surveyheading{Algorithms}
  & \surveyheading{GPU\\Runtimes} & \surveyheading{Architectural\\Support}
  & \surveyheading{Workload\\Analysis}
  & \surveyheading{Controlled\\Profiling} \\
\midrule
General Overviews & Chen and Wang~\cite{ref03}, Wu et al.~\cite{ref05}, Bao et al.~\cite{ref06}, Fei et al.~\cite{fei20253dgs_survey} & \primary & \partialscope & \partialscope & \partialscope & \outside \\
Compression & Bagdasarian et al.~\cite{ref08}, Ali et al.~\cite{ref09}, Youn et al.~\cite{ref10} & \primary & \partialscope & \outside & \partialscope & \partialscope \\
Applications & He et al.~\cite{ref14}, Tosi et al.~\cite{ref11}, Zhu et al.~\cite{ref12}, Qiu et al.~\cite{ref13}, Dahal et al.~\cite{dahalSurvey} & \primary & \partialscope & \partialscope & \primary & \primary \\
Quality Assessment & Liu et al.~\cite{liuQuality26} & \partialscope & \outside & \outside & \outside & \outside \\
Hardware Accelerators & Peng et al.~\cite{ref16} & \partialscope & \partialscope & \primary & \primary & \partialscope \\
Ours & \textemdash{} & \primary & \primary & \primary & \primary & \primary \\
\bottomrule
\end{tabularx}
\par\smallskip\raggedright
Each row aggregates the strongest coverage within its cited group. In multi-paper rows, different columns may reflect different reviews. \primary{} systematic analysis or evaluation, \partialscope{} brief discussion or limited scope, \outside{} not covered. Workload Analysis examines stage-level bottlenecks or workload-count trade-offs. Controlled Profiling denotes matched-input measurements of execution stages, memory, or latency conducted by the survey authors.
\end{table*}
Existing surveys reflect these separate scopes rather than connecting them (Table~\ref{tab:survey-comparison}). Prior reviews examine general rendering pipelines, compression methods, domain applications in SLAM, robotics, extended reality, and reconstruction, visual quality assessment, or dedicated accelerators~\cite{ref03,ref05,ref06,fei20253dgs_survey,ref08,ref09,ref10,ref14,ref11,ref12,ref13,dahalSurvey,liuQuality26,ref16}. Although several reviews include direct measurements of compression, SLAM, reconstruction, or isolated rendering stages, none connects algorithmic representations, GPU runtimes, and architectural designs along a shared workload path paired with matched-input profiling. Auditing 59 representative efficiency studies in our corpus confirms why literature aggregation alone cannot establish end-to-end causality: while 98.3\% jointly report visual quality and end-to-end performance, only 50.8\% originally pair them with intermediate workload counts and stage timing and only 49.2\% with workload and memory or traffic evidence; even after supplementing missing measurements via our GSPROF reproductions, joint coverage reaches only 54.2\% and 57.6\%.

\begin{figure*}[!t]
  \centering
  \includegraphics[width=\textwidth]{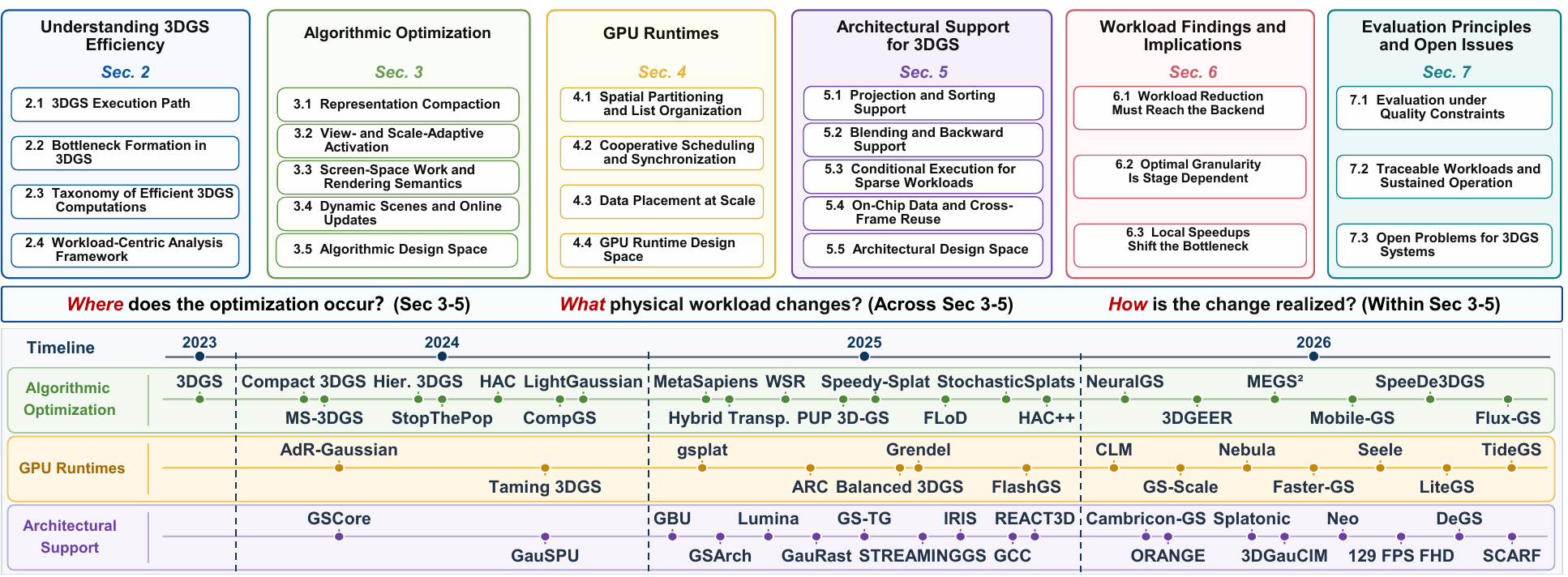}
  \caption{Organization of the survey and timeline of efficient 3DGS systems. The upper panels outline Sections~2--7 and show how the survey classifies optimizations by where they intervene, what workload they change, and how they achieve that change. The lower timeline shows representative studies in Algorithmic Optimization, GPU Runtimes, and Architectural Support from 2023 to 2026.}
  \label{fig:overview}
\end{figure*}

To compare algorithms, GPU runtimes, and architectural support on a consistent physical basis, this survey anchors prior research on the shared Projection, Sorting, and Blending rendering path while extending the analysis to training and online updates. Along the Forward path, our workload-centric framework traces five physical workload counts from resident Gaussians through active Gaussians, Gaussian-tile associations, and pixel evaluations to final blended contributions. Across both rendering and updates, we analyze each study through three linked questions: where an optimization first intervenes, what physical computation or data movement it alters, and how that change is realized through mechanisms such as compact encoding, work pruning, task or data reorganization, result reuse, and specialized hardware mapping. Comparisons retain the task, quality setting, platform, and timed operations so that direct numerical comparisons remain restricted to aligned conditions, while unmatched results support mechanism and trend analysis. Where published studies leave causal behavior unresolved, we complement literature synthesis with controlled GPU profiling under matched inputs, camera trajectories, and visual-quality constraints using GSPROF, an open-source profiling toolkit we developed for this survey.

Synthesizing the full stack through this taxonomy and controlled profiling yields three concrete observations that qualify common efficiency assumptions. First, reducing stored bytes or geometric workload counts translates into end-to-end speedup only when downstream savings outweigh attribute decoding and intermediate materialization; otherwise, data staging can offset geometric reductions, invert local kernel versus end-to-end rankings, or produce sharply different frame costs depending on whether decoding occurs at load time or on every frame. Second, different stages favor different granularities, and static screen-space policies such as fixed tile sizes or fixed culling thresholds do not preserve a single optimum across scenes and resolutions; aggressive culling can violate visual-quality bounds or lose effectiveness at higher resolutions, while average frame rate and even P95 latency can still mask P99 frame-budget violations along interactive camera trajectories. Third, accelerating Forward rasterization shifts limiting costs during training and sustained operation toward parameter updates, tiered memory transfers, and cross-frame recomputation, while unaligned rendering defaults such as near-plane clipping can introduce large image discrepancies that confound cross-backend comparisons. Building on these findings, we formulate quality-constrained evaluation principles and identify open problems in shared workload interfaces, adaptive execution, and programmable hardware specialization. Fig.~\ref{fig:overview} summarizes the survey organization and representative milestones across all three layers.

Section~\ref{sec:fundamentals} describes the 3DGS execution path and defines the workload counts. Sections~\ref{sec:algorithm}--\ref{sec:architecture} examine algorithms, GPU runtimes, and architectural support, respectively. Section~\ref{sec:synthesis} synthesizes the cross-layer findings. Section~\ref{sec:evaluation} presents evaluation principles and open issues, and Section~\ref{sec:conclusion} concludes.

\section{Understanding 3DGS Efficiency}
\label{sec:understanding}
\label{sec:fundamentals}

The computation, storage, and data-movement costs of 3DGS arise when an explicit scene representation is converted into view-dependent rendering work. This section first describes that execution path and explains how its bottlenecks form. It then organizes prior efficiency research and defines the workload vocabulary used throughout the paper.

\subsection{3DGS Execution Path}
\label{sec:execution_path}
\label{sec:3dgs-fundamentals}

A 3D Gaussian stores spatial position, shape, opacity, and appearance. Its center is represented by the mean $\mu_i$ and its 3D shape is described by the covariance $\Sigma_i$. The covariance is commonly parameterized by a scale matrix $S_i$ and rotation matrix $R_i$:
\begin{equation}
\Sigma_i = R_i S_i S_i^T R_i^T.
\end{equation}
Appearance combines opacity and view-dependent color. The original 3DGS~\cite{ref02} represents the latter with spherical harmonics (SH). For a camera view, the Gaussian center and covariance are projected into image space, and the viewing direction determines the SH basis used to evaluate color from the stored coefficients. The screen-space covariance can be summarized as
\begin{equation}
\Sigma_i' = J_i W \Sigma_i W^T J_i^T,
\end{equation}
where $W$ is the camera transform and $J_i$ is the local Jacobian of perspective projection at the Gaussian center. The resulting two-dimensional covariance determines the projected footprint and the tiles and pixels that the Gaussian can affect.

\begin{figure*}[t]
  \centering
  \includegraphics[width=\textwidth]{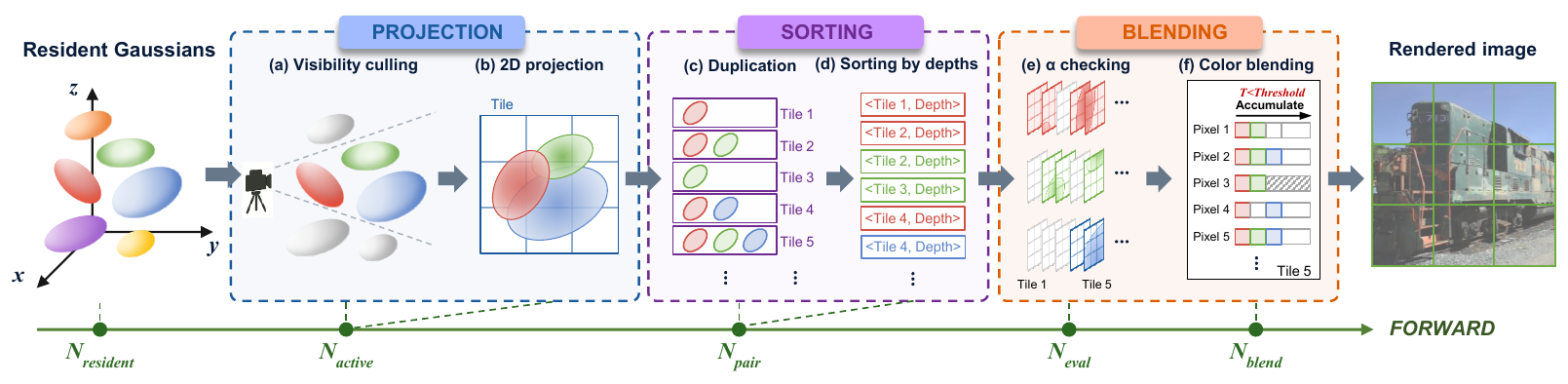}
  \caption{3DGS Forward path with the five counts used in the analysis. Forward rendering proceeds from Gaussians stored in rendering memory through visibility culling, projection, Gaussian--tile association construction, depth ordering, candidate evaluation, and alpha compositing.}
  \label{fig:rendering-path-workload-model}
\end{figure*}

Standard GPU implementations follow the tile-based Forward path in Fig.~\ref{fig:rendering-path-workload-model}. Visibility processing selects Gaussians for the current view, and Projection produces their screen-space footprints and depths. Each footprint is associated with the tiles that it overlaps, constructing explicit Gaussian-tile records. The records are grouped by tile and ordered by depth. Pixel threads then traverse the ordered list of their tile and evaluate Gaussian response, opacity, transmittance, and color contribution.

For pixel $p$, let $\alpha_i(p)$ denote the effective opacity of Gaussian $i$. Assuming a black background (omitting any residual background term), Gaussians covering the same pixel are composed in front-to-back order to form the foreground color:
\begin{equation}
C(p) = \sum_{i} T_i(p)\alpha_i(p)c_i, \qquad T_i(p) = \prod_{j<i}\bigl(1 - \alpha_j(p)\bigr).
\end{equation}
The compositing result is order dependent. Contribution tests can reject weak candidates, and traversal can terminate after accumulated opacity reaches a threshold. Only accepted contributions update pixel color and transmittance.

We refer to the three stages of Forward as Projection, Sorting, and Blending. Projection covers visibility processing, screen-space projection, and footprint construction. Sorting covers Gaussian-tile association construction, grouping by tile, and depth ordering. Blending covers pixel-level Gaussian evaluation and ordered alpha compositing. Throughout this survey, rasterization denotes this pixel-level work within Blending, and association construction belongs to Sorting. Forward denotes the complete Projection--Sorting--Blending path and forms the primary execution spine examined in this survey. Training and online systems extend this path with Backward, which propagates image-space losses to Gaussian parameters and accumulates gradients from many pixels, and Update, which applies parameter and density-control updates~\cite{ref02}.

\subsection{Bottleneck Formation in 3DGS}
\label{sec:bottlenecks}
\label{sec:3dgs-bottlenecks}

Because 3DGS stores scene geometry and appearance explicitly across millions of Gaussians, each described by dozens of parameters, it avoids neural network queries at the cost of reading stored attributes and rebuilding view-dependent work for every frame~\cite{ref02}. Larger scenes expand the Gaussian set, while higher output resolutions or broader projected coverage increase association records and pixel evaluations along the Forward path, raising computation and data access~\cite{ref38}. On desktop GPUs, these view-dependent bursts limit frame time, throughput, and tail-latency stability under interactive frame budgets. Edge devices face tighter memory, bandwidth, and power budgets that can prevent real-time execution. Most inference traffic arises from repeated accesses to Gaussian attributes and from constructing, ordering, and traversing per-frame association records.

Studies across platforms and workloads identify Blending as a major source of inference time~\cite{ref44,ref71,ref95}. Each pixel traverses depth-ordered Gaussians and repeatedly evaluates responses, opacity, and color contributions. Because Gaussians are distributed irregularly in space, projected association counts and useful blending work vary widely across tiles. The resulting imbalance lowers SIMT utilization and degrades locality in attribute accesses~\cite{ref54,ref55,ref85}. Once dedicated hardware or runtime optimization accelerates rasterization, Sorting and its DRAM traffic can emerge as the next major cost~\cite{ref48}. The dominant bottleneck therefore depends on the scene, viewpoint, resolution, platform, and optimizations already applied to each stage.

Training and online operation introduce additional pressures beyond Forward rendering. During Backward, gradients from many pixels converge on shared Gaussians, causing atomic memory conflicts and serialized stalls. Gaussian parameters, gradients, and optimizer states also expand memory footprint as scene scale grows~\cite{ref73}; once these structures exceed GPU capacity, host-device transfers can dominate iteration time~\cite{ref61,ref62}. Dynamic scenes must further update Gaussian geometry and appearance over time, while optimization-based SLAM alternates pose estimation, rendering, gradient propagation, and map updates across consecutive frames~\cite{ref81,ref49}. Changes in viewpoint, occlusion, or scene content can invalidate prior visibility results, tile associations, and depth order, restricting temporal reuse~\cite{ref48,ref49}. Together, these inference and update pressures explain why efficiency research intervenes at multiple points along the execution path.

\subsection{Taxonomy of Efficient 3DGS Computations}
\label{sec:taxonomy}
\label{sec:survey-organization}

Existing efficiency studies fall into three primary design domains. Algorithms modify stored Gaussian representations, the subset of Gaussians activated for a view, or rendering semantics. GPU Runtimes organize association lists, pixel and gradient scheduling, and multi-tier data placement on programmable processors. Architectural Support maps these operations and their data onto specialized compute units, on-chip buffers, and memory subsystems. Because individual systems can span multiple domains, we place each study in Sections~\ref{sec:algorithm}--\ref{sec:architecture} according to its central design contribution and trace its end-to-end effects through the framework below. \supsec{1} describes our literature screening criteria, and \supsec{2} documents the reproducibility protocol and our open-source GSPROF toolkit.

We analyze each study through three linked questions. \emph{Where does the optimization occur?} This identifies the first affected operation along the rendering or update path. \emph{What physical workload changes?} We trace representation size, per-frame rendering work, data movement, and synchronization, alongside retained depth orders, cached images, gradients, optimizer states, or update masks carried into subsequent stages. \emph{How is the change realized?} We identify the underlying engineering mechanism, including pruning or filtering redundant work, compactly encoding attributes, reorganizing execution schedules or data layouts, reusing cached or historical results, and mapping selected operations to specialized hardware.

As summarized in Fig.~\ref{fig:overview}, Sections~\ref{sec:algorithm}--\ref{sec:architecture} apply these three questions consistently across Algorithms, GPU Runtimes, and Architectural Support, spanning representation design, scheduling and data placement, and hardware mapping. Task, resolution, quality setting, rendering defaults, platform, and timing scope jointly define the comparison context. Direct numerical comparisons are restricted to aligned conditions, whereas results from unmatched settings are used only to explain mechanisms and trends within each study's reported conditions.

\subsection{Workload-Centric Analysis Framework}
\label{sec:framework}
\label{sec:workload-analysis}

\begin{figure*}[!t]
  \centering
  \includegraphics[width=\textwidth]{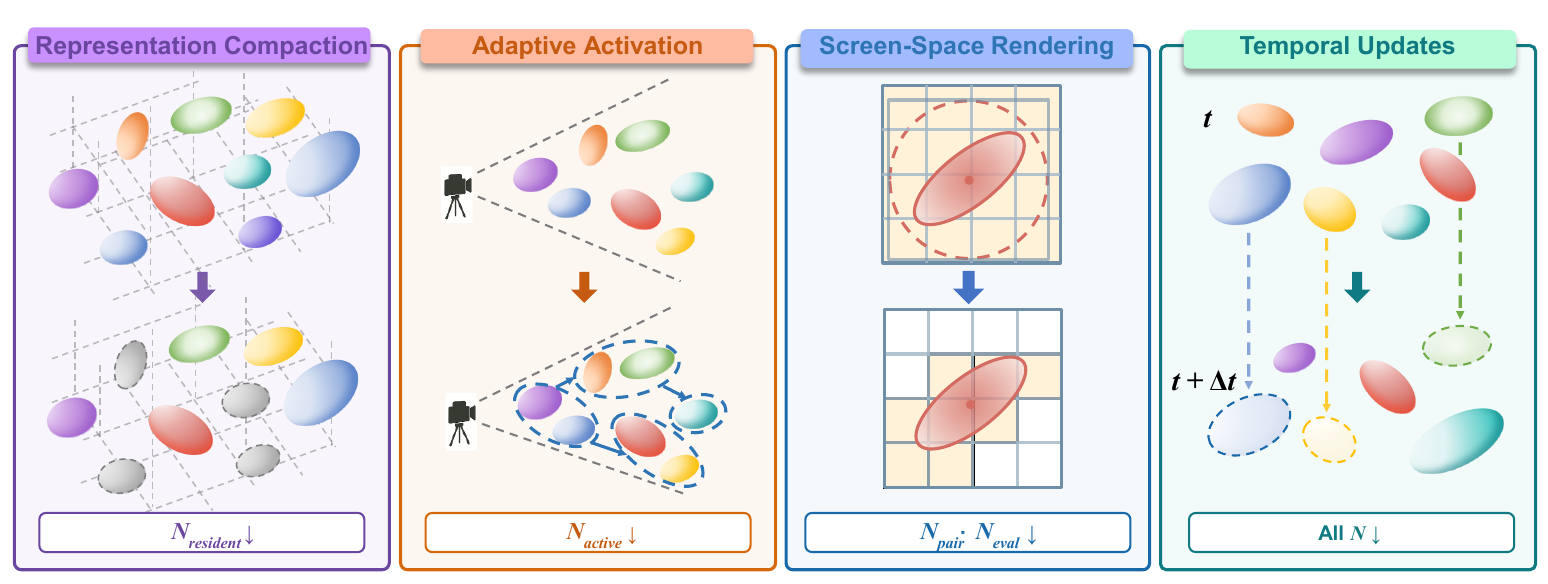}
  \caption{Representative 3DGS optimizations span representation compaction, adaptive activation, screen-space rendering, and temporal updates. In the workload labels, $\downarrow$ means that the indicated Forward counts decrease or stay the same.}
  \label{fig:algorithmic-workload-optimization}
\end{figure*}

To connect isolated optimizations with end-to-end system effects, we characterize the Forward path in Fig.~\ref{fig:rendering-path-workload-model} using five physical workload counts:
\begin{equation*}
N_{\text{resident}} \rightarrow N_{\text{active}} \rightarrow N_{\text{pair}} \rightarrow N_{\text{eval}} \rightarrow N_{\text{blend}}.
\end{equation*}
These counts begin with Gaussians stored in rendering memory and follow view selection, screen-space expansion, pixel evaluation, and compositing. The arrows indicate execution order rather than monotonic reduction, because each stage measures a distinct work item.

At the start of the path, $N_{\text{resident}}$ counts the Gaussians stored in the rendering memory pool and directly accessible to the renderer under test. This pool may reside in device-local, unified, or host-mapped memory, and its memory tier must be reported explicitly. In hierarchical or streamed systems, the complete level-of-detail (LoD) hierarchy may remain on the host, in the cloud, or in external storage; Gaussians outside the measured rendering pool are excluded from $N_{\text{resident}}$. Together with the bytes stored per Gaussian, $N_{\text{resident}}$ determines the memory footprint of stored attributes.

$N_{\text{active}}$ counts Gaussians retained after view selection, visibility culling, and projected-footprint tests. In our controlled profiles, these are the distinct Gaussians producing at least one materialized association. This post-projection count excludes rejected inputs that still incur Projection work.

Projected footprints are expanded into Gaussian-tile association records during Sorting. $N_{\text{pair}}$ counts the association records produced by this expansion. Because a single active Gaussian can overlap multiple tiles, $N_{\text{pair}}$ typically exceeds $N_{\text{active}}$ and governs association writes, depth-sorting work, and subsequent list reads. Depth ordering reorganizes these records without altering their count.

When pixel threads traverse tile lists during Blending, $N_{\text{eval}}$ counts the Gaussian response or contribution tests they execute; a rejected candidate still counts because its test has already run. Only a subset of evaluated candidates reaches compositing. $N_{\text{blend}}$ counts the contributions that pass contribution tests and update pixel color and transmittance. Within the same frame and backend, $N_{\text{blend}} \le N_{\text{eval}}$. With the scene, viewpoint, resolution, and backend semantics held fixed, ideal lossless prefiltering leaves $N_{\text{blend}}$ unchanged, and the pixel loop reaches the lower bound $N_{\text{eval}} = N_{\text{blend}}$ only when every non-contributing candidate is rejected before evaluation; the cost of the prefilter must be accounted for separately.

The ratios between successive counts vary with scene content, viewpoint, resolution, tile configuration, culling policy, and backend semantics, so reducing an upstream count need not lower downstream counts proportionally. Moreover, while these five counts capture stored Gaussians and Forward operations, end-to-end system cost also depends on work per item, attribute decoding and materialization, temporary buffers, transfers, synchronization, cached results, and optimizer states. Backward gradient accumulation and Update operations add further cost. Reducing a workload count therefore saves time or resources only when the implementation eliminates actual computation, memory accesses, synchronization events, or updates without introducing offsetting overhead.

The Forward counts also help interpret costs outside rendering. The ratio $N_{\text{blend}}/N_{\text{active}}$ estimates average contribution fan-in per active Gaussian and indicates potential Backward atomic-write contention, with actual contention depending on how contributions are distributed and scheduled. For a fixed parameter layout, optimizer-state capacity and full Update computation grow with $N_{\text{resident}}$, while selective updates can restrict the set of parameter writes without freeing resident optimizer state. Dynamic 4DGS adds a per-frame deformation step upstream of Projection. Skipping deformation for static Gaussians cuts this work even when all five Forward counts stay the same. The five workload counts instantiate our framework for conventional tile-based Gaussian rasterization. For alternative rendering paths such as ray tracing and sorting-free compositing, the same analysis follows the corresponding candidate-generation, evaluation, and accumulation operations using path-specific intermediate work objects.

\section{Algorithmic Optimization}
\label{sec:algo}
\label{sec:algorithm}

\begin{table*}[!t]
\caption{Representative Algorithmic Methods and Workload Effects}
\label{tab:algorithm-overview}
\centering
\scriptsize
\setlength{\tabcolsep}{2.5pt}
\renewcommand{\arraystretch}{1.15}
\begin{tabularx}{\textwidth}{@{}>{\raggedright\arraybackslash}m{0.133\textwidth}>{\raggedright\arraybackslash}p{0.24\textwidth}>{\raggedright\arraybackslash}p{0.232\textwidth}>{\raggedright\arraybackslash}p{0.188\textwidth}Y@{}}
\toprule
Category & Method & Mechanism & Target & Overhead / Bound \\
\midrule
\multirow{13}{=}{Compaction} & Reducing Memory Footprint~\cite{ref18} & Sparse pruning + quantization & $\Nresident\downarrow$, Bytes$\downarrow$ & Decode, Quality \\
 & LightGaussian~\cite{ref31} & Pruning + SH distillation & $\Nresident\downarrow$, Bytes$\downarrow$ & Fine-tune, Quality \\
 & Trimming the Fat~\cite{ref33} & Iterative importance pruning & $\Nresident\downarrow$, $\Neval\downarrow$ & Fine-tune, Quality \\
 & MEGS2~\cite{ref21} & Spherical Gaussians + unified pruning & $\Nresident\downarrow$, Bytes$\downarrow$ & Fine-tune, Quality \\
 & PUP 3D-GS~\cite{ref76} & Reconstruction sensitivity pruning & $\Nresident\downarrow$ & Quality \\
 & Provable Pruning via Coresets~\cite{ref24} & Coreset pruning & $\Nresident\downarrow$ & Theory, Quality \\
 & Flux-GS~\cite{ref23} & Energy aggregation + compaction & $\Nresident\downarrow$, Bytes$\downarrow$ & Thermals \\
\addlinespace[2pt]
 & Compressed 3DGS~\cite{ref25} & Pruning + entropy coding & $\Nresident\downarrow$, Bytes$\downarrow$ & Decode, \mbox{Co-design} \\
 & Compact 3DGS~\cite{ref29} & Masking + vector quantization & $\Nresident\downarrow$, Bytes$\downarrow$ & Decode, Quality \\
 & CompGS~\cite{compgsVQ} & Shared attribute codebooks & Bytes$\downarrow$ & Decode \\
 & HAC~\cite{ref32} & Hash-grid context entropy coding & Bytes$\downarrow$ & Decode \\
 & HAC++~\cite{ref50} & Enhanced context entropy coding & Bytes$\downarrow$ & Decode \\
 & CodecGS~\cite{ref51} & Feature planes + video coding & Bytes$\downarrow$ & Decode \\
\midrule
\multirow{4}{=}{Adaptive activation} & Hierarchical 3DGS~\cite{ref17} & LoD hierarchy + streaming & $\Nactive\downarrow$ & Selection, Memory \\
 & MetaSapiens~\cite{ref19} & Guided pruning + foveation & $\Nactive\downarrow$, $\Neval\downarrow$ & Quality, \mbox{Co-design} \\
 & FLoD~\cite{ref28} & Selectable detail levels & $\Nactive\downarrow$ & Selection, Quality \\
 & Multi-Scale 3DGS~\cite{ref30} & Multiscale coverage filter & $\Nactive\downarrow$, $\Neval\downarrow$ & Selection, Memory \\
\midrule
\multirow{7}{=}{Screen-space work} & Easily Integrable Improvements~\cite{ref26} & Culling + adaptive tiling & $\Npair\downarrow$, $\Neval\downarrow$ & Selection \\
 & StopThePop~\cite{ref27} & Hierarchical Sorting + culling & $\Npair\downarrow$, order error & Quality, Memory \\
 & Hybrid Transparency~\cite{ref58} & Partial order + unordered tail & Sort$\downarrow$, blend rule & Semantics, Quality \\
 & StochasticSplats~\cite{ref60} & Stochastic rasterization & Sort$\downarrow$, blend rule & Semantics, Quality \\
 & Speedy-Splat~\cite{ref77} & Sparse Gaussian + pixel culling & $\Nactive\downarrow$, $\Neval\downarrow$ & Quality \\
 & Mobile-GS~\cite{ref34} & Unordered blending + decoding & $\Nresident\downarrow$, Sort$\downarrow$ & Decode, Thermals \\
 & 3DGEER~\cite{ref36} & Exact ray rendering & $\Neval\downarrow$, ray candidates & Semantics \\
\midrule
\multirow{3}{=}{Dynamic updates} & SpeeDe3DGS~\cite{ref20} & Temporal pruning + motion grouping & Deform$\downarrow$ & Selection, Quality \\
 & RTGS~\cite{ref22} & Pruning + pixel downsampling & $\Neval\downarrow$, Update$\downarrow$ & Co-design \\
 & Mobile Avatars~\cite{ref35} & Blendshape pruning + light decoding & Bytes$\downarrow$, Deform$\downarrow$ & Decode, Thermals \\
\bottomrule
\end{tabularx}
\par\smallskip
\parbox{\textwidth}{\footnotesize $\downarrow$ denotes a targeted reduction. Symbols follow Section~\ref{sec:fundamentals}. Overhead / Bound lists added costs, quality constraints, or evaluation scope.}
\end{table*}

Algorithmic efficiency in 3DGS is commonly judged through representation-level metrics such as file size or Gaussian count, yet comparable reductions in stored data routinely yield starkly different rendering speeds. A purely technique-oriented grouping obscures why this divergence occurs. Viewed through our workload-centric framework, algorithmic methods differ primarily in where they enter the execution path and whether their upstream reductions propagate from $N_{\text{resident}}$ to $N_{\text{blend}}$ once decoding, selection, and update costs are accounted for (Fig.~\ref{fig:algorithmic-workload-optimization}, Table~\ref{tab:algorithm-overview}). Following this causal chain separates optimizations that merely compact stored state from those that genuinely relieve downstream Sorting, Blending, and memory traffic.

\subsection{Representation Compaction}
\label{sec:compaction}

Representation compaction lowers the number of Gaussians stored in rendering memory, $N_{\text{resident}}$, or the storage required per Gaussian. Pruning removes Gaussians that contribute little to visual reconstruction. Reducing the Memory Footprint~\cite{ref18} combines sparse regularization, pruning, and quantization, and LightGaussian~\cite{ref31} pairs importance pruning with SH distillation and vector quantization. Trimming the Fat~\cite{ref33} and MEGS2~\cite{ref21} refine pruning criteria and spherical representations. PUP 3D-GS~\cite{ref76} scores Gaussians by reconstruction sensitivity, while Speedy-Splat~\cite{ref77} reduces scoring storage so pruning can run during training. Coreset pruning~\cite{ref24} provides resolution-dependent approximation guarantees under explicit assumptions on rendering queries and transmittance. Discarding stored Gaussians lowers $N_{\text{active}}$ and later work only in views that would otherwise render them.

\begin{figure*}[!t]
  \centering
  \includegraphics[width=\textwidth]{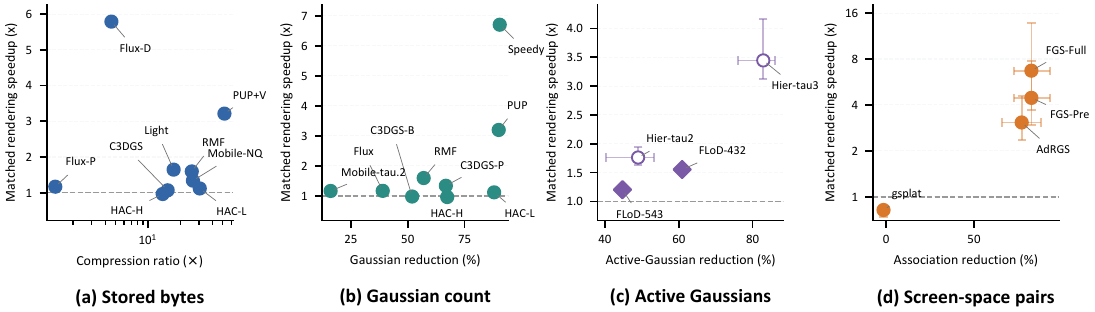}
  \caption{Rendering speedup versus (a) storage compression ratio and percentage reductions in (b) Gaussians, (c) active Gaussians, and (d) Gaussian--tile associations, matched within each source. Similar association reductions can yield different speedups. Panel (d) uses a base-2 logarithmic vertical scale, while panels (a)--(c) use linear vertical scales. Bars span minima to maxima, and hollow markers indicate contextual comparisons. \supsec{2} identifies the plot labels. Section~\ref{sec:runtime} discusses the GPU runtimes in panel (d).}
  \label{fig:compression-speed}
\end{figure*}

Attribute compression shrinks stored size without necessarily reducing $N_{\text{resident}}$. Compact 3DGS~\cite{ref29}, CompGS~\cite{compgsVQ}, Compressed 3DGS~\cite{ref25}, and HAC~\cite{ref32} compact color and SH data through shared vector codebooks, quantization, parameterization, and entropy coding. CodecGS~\cite{ref51}, HAC++~\cite{ref50}, TC-GS~\cite{ref52}, and NeuralGS~\cite{ref53} rely on video or entropy decoding, feature-plane recovery, hash lookups, or MLP inference to reconstruct attributes. Yang et al.~\cite{colorFields26} take a different route by learning a neural view-dependent color field, which lies outside the fixed third-order SH evaluation examined in Section~\ref{sec:materialization}. For compressed representations, rendering speed depends on attribute recovery and resident memory expansion, so disk size, decoded memory footprint, Gaussian count, and decoding time must be measured separately. Flux-GS~\cite{ref23} couples compaction with mobile rendering, where memory bandwidth and sustained operation also constrain performance. Fig.~\ref{fig:compression-speed}(a)--(b) shows that similar reductions in stored bytes or Gaussian count still yield different rendering speedups within matched sources.

\subsection{View- and Scale-Adaptive Activation}
\label{sec:activation}

Viewpoint distance and output resolution determine which part of a scene requires fine detail. Adaptive activation retains a complete multiresolution representation and exposes a smaller selected subset to Projection for each frame. Hierarchical 3DGS~\cite{ref17} combines a Gaussian hierarchy with CPU--GPU streaming to select LoD Gaussians for the current view. FLoD~\cite{ref28} exposes selectable detail levels for customizable budgets, and Multi-Scale 3DGS~\cite{ref30} filters Gaussians by projected screen coverage and resolution, complementing scale-aware anti-aliasing filters such as Mip-Splatting~\cite{mipSplatting24} that stabilize rendering across sampling rates.

Perceptual importance can also guide selection. MetaSapiens~\cite{ref19} combines efficiency-aware pruning with foveated rendering to activate fewer Gaussians in peripheral regions. Lowering $N_{\text{active}}$ saves Sorting and Blending work according to the screen-space footprints of the omitted Gaussians, while adding recurring costs for hierarchy traversal, per-frame selection, prefetching, and level transitions. Fig.~\ref{fig:compression-speed}(c) relates active-set reduction to rendering speedup across matched sources, though isolating the saved footprint work from selection overhead requires stage-level timing.

\subsection{Screen-Space Work and Rendering Semantics}
\label{sec:screenspace}

Screen-space methods act after Projection to shrink candidate lists or reduce per-candidate evaluation cost. Easily Integrable Improvements~\cite{ref26} combines tighter culling, adaptive tiling, and faster evaluation over an unchanged representation. Speedy-Splat~\cite{ref77} pairs tight screen-space bounding with sparse Gaussian and pixel processing to skip low-contribution work. Tighter coverage directly lowers $N_{\text{pair}}$ entering Sorting and tile traversal, whereas faster response evaluation reduces compute cost per candidate without necessarily lowering $N_{\text{eval}}$. Fig.~\ref{fig:compression-speed}(d) relates association reductions to rendering speedup across matched implementations, including the GPU runtime methods examined in Section~\ref{sec:runtime}, showing that similar reductions in $N_{\text{pair}}$ can produce different speedups.

Changing depth ordering or alpha compositing affects both workload and image semantics. StopThePop~\cite{ref27} uses hierarchical sorting and tighter culling to improve view consistency, and 3DGEER~\cite{ref36} constructs candidates for exact ray-based volume rendering across generic cameras. Hybrid Transparency~\cite{ref58}, WSR~\cite{ref59}, and StochasticSplats~\cite{ref60} relax strict front-to-back ordering through partial ordering, weighted accumulation, or stochastic sampling to avoid full depth sorting. Mobile-GS~\cite{ref34} pairs unordered blending with compact attributes and lightweight decoding for mobile devices, while Koriyama and Watanabe~\cite{blueNoise26} apply blue-noise stochastic sampling to dynamic projection mapping, altering the spatial and temporal distribution of rendering error. Because these rules change the rendered output itself, their speedups must be assessed alongside image fidelity, view consistency, and the compositing rule used.

\subsection{Dynamic Scenes and Online Updates}
\label{sec:dynamic}
\label{sec:dynamic-updates}

Dynamic rendering and online reconstruction modify Gaussian geometry, appearance, or optimizer states across consecutive frames. In 4D Gaussian Splatting~\cite{ref81}, a learned deformation field maps canonical Gaussians to time-dependent positions and shapes before Projection. Deformation can therefore dominate a frame even when its downstream Forward workload is comparable to that of a static scene. For dynamic scenes, SpeeDe3DGS~\cite{ref20} uses temporal pruning and motion grouping to avoid deforming static or redundant Gaussians every frame, and Mobile Avatars~\cite{ref35} compresses local blendshapes with lightweight decoding for mobile devices. Both cut per-frame deformation and rendering work, while adding motion-group lookups or blendshape decoding to the rendering path. Skipping deformation on static Gaussians targets the upstream work described in Section~\ref{sec:framework}, with any savings in the five Forward counts depending on the resulting view-selected set and projected footprints.

In training and online mapping, efficiency also depends on limiting redundant Backward and Update work. RTGS~\cite{ref22}, revisited in Section~\ref{sec:conditional-execution}, reuses keyframe and gradient information to drive dynamic pixel downsampling and Gaussian pruning, reducing both Forward evaluation and parameter-update work across frames. The resulting Update savings depend on which parameters are written, while optimizer state may still be retained for the entire resident set.

\subsection{Algorithmic Design Space}
\label{sec:algo_space}

The runtime payoff of an algorithmic intervention depends on how often it executes and how much downstream work it removes. Offline pruning and attribute encoding spread their cost over many views of a static scene, although compressed attributes can still incur decoding during rendering. Adaptive activation and screen-space culling adapt to current footprints but execute on every frame, whereas temporal reuse avoids repeated work only until motion or appearance changes require recomputation. Reducing storage volume or upstream counts therefore improves frame time only when the eliminated Sorting, Blending, and memory traffic exceed the added cost of decoding, hierarchy traversal, selection, or deformation lookup. When compaction, aggressive culling, or relaxed transparency alters the rendered signal, those savings must also stay within explicit image-quality and view-consistency bounds.

\section{GPU Runtimes}
\label{sec:runtime}

\begin{figure}[!t]
  \centering
  \includegraphics[width=\columnwidth]{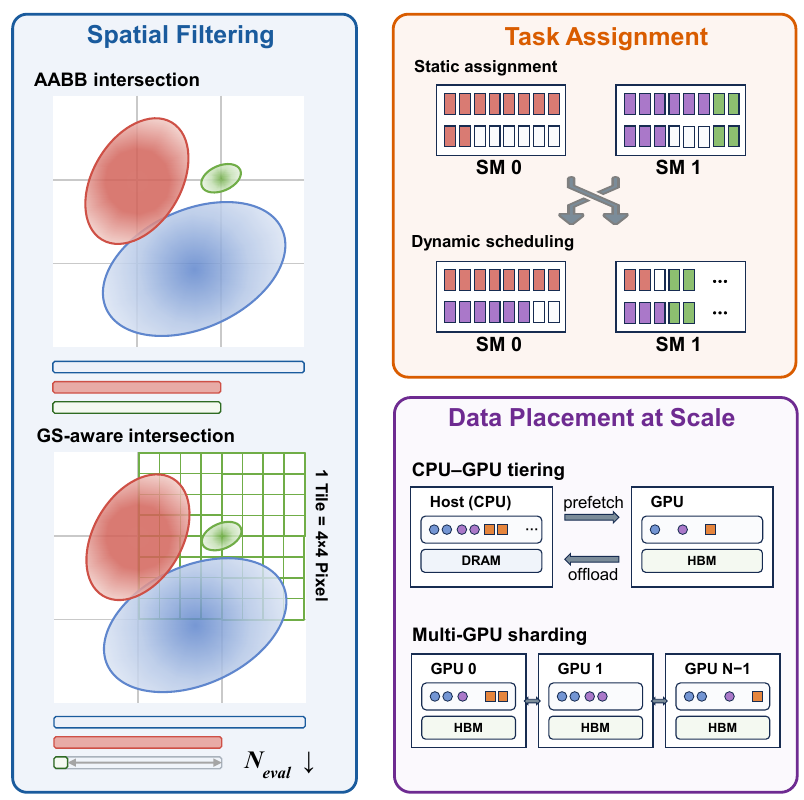}
  \caption{GPU runtime filtering, scheduling, and data placement. Gaussian-aware intersection reduces invalid evaluations, dynamic scheduling redistributes uneven tasks, and CPU--GPU tiering or multi-GPU sharding distributes Gaussian parameters and optimizer data.}
  \label{fig:runtime-map}
\end{figure}

Even when the scene representation and active Gaussian set remain fixed, GPU implementations routinely exhibit wide gaps in latency, memory footprint, and DRAM traffic. Treating runtime techniques as isolated kernel optimizations obscures the structural cause of this spread: on programmable SIMT hardware, irregular screen-space footprints turn every spatial partitioning, scheduling, and placement decision into a trade-off between intermediate work volume, parallel utilization, and data movement (Fig.~\ref{fig:runtime-map}, Table~\ref{tab:runtime-overview}). Examining GPU runtimes through our workload framework distinguishes designs that shrink intermediate associations and pixel tests from those that reorganize execution and memory placement over an unchanged workload.

\begin{table*}[!tp]
\caption{Representative GPU Runtime Methods and Workload Effects}
\label{tab:runtime-overview}
\centering
\scriptsize
\setlength{\tabcolsep}{2.5pt}
\renewcommand{\arraystretch}{1.15}
\begin{tabularx}{\textwidth}{@{}>{\raggedright\arraybackslash}m{0.163\textwidth}>{\raggedright\arraybackslash}p{0.17\textwidth}>{\raggedright\arraybackslash}p{0.075\textwidth}>{\raggedright\arraybackslash}p{0.27\textwidth}Y@{}}
\toprule
Category & Method & Scope & Mechanism & Workload Effect \\
\midrule
\multirow{4}{=}{Spatial partitioning} & FlashGS~\cite{ref38} & Forward & Exact overlap + compact lists & $\Npair\downarrow$, Sort$\downarrow$ \\
 & AdR-Gaussian~\cite{ref100} & Forward & Adaptive tight AABBs & $\Npair\downarrow$, Sort$\downarrow$ \\
 & Seele~\cite{ref39} & Forward & Preprocessing + contribution tests & $\Neval\downarrow$ \\
 & HiGS~\cite{ref55} & Forward & Macro-tile Sorting + fine rasterization & Sort$\downarrow$, $\Neval\downarrow$ \\
\midrule
\multirow{5}{=}{Cooperative scheduling} & Balanced 3DGS~\cite{ref54} & Forward & Dynamic tile scheduling + warp rendering & Load bal. \\
 & ARC~\cite{ref56} & Backward & Warp reduction + cache cooperation & Atomics$\downarrow$ \\
 & Taming 3DGS~\cite{ref57} & Backward & Gaussian-centric gradient reduction & Atomics$\downarrow$ \\
 & Faster-GS~\cite{ref96} & Backward & Gaussian reduction + spatial layout & Atomics$\downarrow$, Locality \\
 & Local-GS~\cite{ref66} & Forward & Parameter hoisting + warp culling & Reuse, $\Neval\downarrow$ \\
\midrule
\multirow{8}{=}{Data placement} & gsplat~\cite{ref37} & Training & Memory-aware CUDA operators & VRAM$\downarrow$ \\
 & CLM~\cite{ref61} & Training & Host offload + microbatch prefetch & VRAM$\downarrow$, Transfer overlap \\
 & GS-Scale~\cite{ref62} & Training & Selective offload + deferred updates & VRAM$\downarrow$, Transfer overlap \\
 & TideGS~\cite{ref98} & Training & SSD prefetch + writeback & VRAM$\downarrow$, Transfer overlap \\
 & Grendel~\cite{ref63} & Training & Sharding + sparse all-to-all & Comm.$\downarrow$ \\
 & LiteGS~\cite{ref64} & Training & Spatial grouping + fused operators & Locality \\
 & LOBE-GS~\cite{ref65} & Training & Visibility cropping + balanced partitioning & Load bal., Locality \\
 & Nebula~\cite{ref99} & Remote & Cloud LoD + client caching & Reuse, Comm.$\downarrow$ \\
\bottomrule
\end{tabularx}
\par\smallskip
\parbox{\textwidth}{\footnotesize $\downarrow$ denotes a targeted reduction. Load bal.\ means load balance. Comm.\ means communication. Scope denotes the stage or execution context targeted by the listed mechanism. Workload Effect reports the primary targeted runtime effect. Section~\ref{sec:runtime_space} and Fig.~\ref{fig:runtime-workload-execution-synthesis} analyze net speedup and trade-offs.}
\end{table*}

\subsection{Spatial Partitioning and List Organization}
\label{sec:runtime_spatial}

\begin{figure*}[!t]
  \centering
  \includegraphics[width=\textwidth]{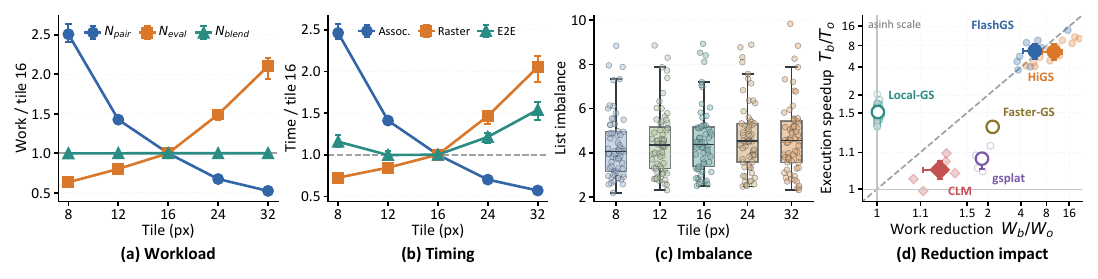}
  \vspace{-20pt}
  \caption{Runtime evidence under controlled tile size changes and matched source conditions. Panels (a)--(c) show workload, timing, and tile list imbalance, where list imbalance in (c) measures each view's P99-to-mean tile list length ratio. The sweep uses 16 views each of Bicycle, Bonsai, Truck, and DrJohnson at width 800, with 30K-iteration checkpoints and tile widths of 8, 12, 16, 24, and 32 pixels. In (b), Assoc. measures association processing within Sorting and Raster measures pixel rasterization within Blending, following Section~\ref{sec:3dgs-fundamentals}. Panel (d) relates workload reduction to speedup under matched conditions within each source. Both axes use $\operatorname{asinh}((r-1)/0.05)$ with ticks labelled by actual ratios $r$ to resolve subtle ratios near $1\times$ alongside speedups up to the $16\times$ tick. The Faster-GS point compares the official testbed's basis and optimized rasterizers under matched inputs and times only the complete Forward pass.}
  \label{fig:runtime-workload-execution-synthesis}
\end{figure*}

Spatial partitioning groups projected Gaussians into per-tile lists that pixel threads traverse during Blending. Tile granularity creates a direct trade-off across stages. Coarse tiles generate fewer lists and scheduling tasks and can improve attribute reuse, yet they pull more Gaussians outside any individual pixel's footprint into the traversal loop. Fine tiles reject more non-overlapping candidates before pixel evaluation, at the cost of expanding association lists and sorting traffic. Controlled profiling in Fig.~\ref{fig:runtime-workload-execution-synthesis}(a)--(b) makes this coupling explicit: shrinking tile width raises $N_{\text{pair}}$ and association-path time within Sorting, whereas enlarging tiles raises $N_{\text{eval}}$ and rasterization time within Blending while leaving $N_{\text{blend}}$ nearly constant. Median normalized Forward latency therefore reaches its minimum between the tested extremes. At the same time, Fig.~\ref{fig:runtime-workload-execution-synthesis}(c) shows that severe tile-list imbalance persists across every tile size, indicating that uniform tiling alone cannot equalize execution load.

Runtime designs relieve this trade-off either by decoupling granularity across stages or by tightening intersection tests before association records are emitted. HiGS~\cite{ref55} uses macro tiles for partitioning and depth sorting, then refines lists into smaller tiles for pixel rasterization to curb invalid evaluations. Seele~\cite{ref39} combines view-dependent preprocessing with fine-grained contribution checks during rasterization to cut wasted pixel work on mobile GPUs. Taking a geometric route, AdR-Gaussian~\cite{ref100} replaces loose circular bounds with tight axis-aligned bounding boxes, while FlashGS~\cite{ref38} performs exact Gaussian-tile intersection testing. Both shrink $N_{\text{pair}}$ directly, reducing Sorting input, intermediate list traffic, and downstream traversal length.

CUDA tile rasterization builds Gaussian--tile lists, sorts them by depth, and uses pixel threads to traverse each tile. Hardware ray-tracing designs such as 3DGRT~\cite{gaussianRayTracing24} instead replace list construction and radix sorting with BVH traversal and proxy-geometry intersection, resolving contribution order along each ray while trading sorting traffic for traversal divergence and ray-coherence sensitivity. Hybrid pipelines such as 3DGUT~\cite{gaussianUnscented25} retain rasterization efficiency for primary rays under distorted cameras while invoking ray tracing for secondary effects, and 3DGEER~\cite{ref36} demonstrates exact ray-based evaluation for generic cameras using its own association scheme. Another route maps splats to fixed-function graphics operations. RayGS~\cite{ref68} uses support quads, and DHR~\cite{ref67} and VR-Pipe~\cite{ref94} illustrate opportunities in blending and fragment processing. Implementations using Vulkan, WebGPU, or Metal can exploit rasterization and blending units exposed by the API. Depth ordering can require synchronization across compute and graphics passes, and subgroup capabilities vary across devices.

\subsection{Cooperative Scheduling and Synchronization}
\label{sec:runtime_sched}

Because tile-list imbalance persists across tile sizes, static mappings from tiles to thread blocks leave execution units idle even when $N_{\text{pair}}$ and $N_{\text{eval}}$ are unchanged. Uneven footprint coverage and early transmittance termination further introduce warp divergence inside each tile. Balanced 3DGS~\cite{ref54} mitigates this skew through dynamic tile assignment paired with Gaussian-parallel rendering across lanes within a warp. Local-GS~\cite{ref66} targets intra-tile execution efficiency by hoisting shared Gaussian parameters, culling inactive candidates at warp granularity, and coordinating coherent pixel blending to increase register and cache reuse.

In Backward, the execution bottleneck shifts from divergent reads to contested writes as gradients from many pixels converge onto shared Gaussians. Pixel-parallel implementations rely on global atomic additions that serialize updates on heavily covered Gaussians. ARC, originally introduced as DISTWAR~\cite{ref56}, aggregates gradients at the warp level and coordinates atomics at the L2 cache to suppress global contention. Taming 3DGS~\cite{ref57} restructures Backward around Gaussians so threads combine gradients locally before writing them back to memory. Faster-GS~\cite{ref96} pairs a refined Gaussian-centric Backward pass with a locality-preserving parameter layout that keeps spatially adjacent Gaussians close in memory. Although these methods compute the same gradients, reducing global atomic traffic and regularizing memory accesses yields substantial iteration speedups.

\subsection{Data Placement at Scale}
\label{sec:runtime_placement}
\label{sec:data-placement}

As scenes scale to millions or billions of Gaussians, parameters, gradients, optimizer states, and rasterization intermediates collectively strain or exceed single-GPU memory capacity. Within a single GPU, gsplat~\cite{ref37} trims allocation overhead through memory-aware CUDA and PyTorch operators. Once the working set outgrows device memory, runtimes tier storage across the host hierarchy. CLM~\cite{ref61} retains compact geometric data needed for visibility culling on the GPU while offloading appearance attributes and optimizer arrays to host DRAM, hiding transfer latency through microbatching, prefetching, and caching. GS-Scale~\cite{ref62} coordinates host-device transfers via parameter forwarding and deferred updates to avoid synchronous stalls, and TideGS~\cite{ref98} extends this hierarchy to SSD storage by streaming active working sets onto the GPU while overlapping prefetch, computation, and writeback.

Multi-GPU and distributed systems partition data and computation across devices, making cross-node communication a central cost. Grendel~\cite{ref63} partitions parameters by Gaussian and rasterization by pixel region, using sparse all-to-all exchanges so each GPU receives only the Gaussians overlapping its assigned pixels. Bridging spatial organization and large-scene memory placement, LiteGS~\cite{ref64} and LOBE-GS~\cite{ref65} pair spatial partitioning and visibility cropping with locality-aware block execution and load balancing to curb irrelevant work across partitions. In interactive remote settings, Nebula~\cite{ref99}, revisited in Section~\ref{sec:blending-backward-support}, splits execution across cloud and client by running level-of-detail selection on the server and transmitting only newly visible Gaussians while caching persistent ones on the client. Across all tiered, sharded, and streamed placements, $N_{\text{resident}}$ counts strictly the Gaussians residing in the measured rendering pool, following the memory-tier rule established in Section~\ref{sec:framework}.

\subsection{GPU Runtime Design Space}
\label{sec:runtime_space}

Spatial partitioning, cooperative scheduling, and data placement interact tightly on programmable GPUs. Shrinking tile size trims $N_{\text{eval}}$ inside Blending while expanding association lists and temporary buffers inside Sorting. Reorganizing Backward around Gaussians relieves atomic contention while altering memory-access patterns across parameter arrays. Offloading optimizer states expands scene capacity while exposing host-device transfer stalls whenever compute outpaces prefetching.

\begin{table*}[!tp]
\caption{Representative 3DGS Hardware Designs and Reported Metrics}
\label{tab:accelerators}
\centering
\scriptsize
\setlength{\tabcolsep}{2.0pt}
\renewcommand{\arraystretch}{1.10}
\begin{tabularx}{\textwidth}{@{}>{\raggedright\arraybackslash}m{0.142\textwidth}>{\raggedright\arraybackslash}p{0.15\textwidth}>{\raggedright\arraybackslash}p{0.103\textwidth}>{\raggedright\arraybackslash}p{0.245\textwidth}>{\raggedright\arraybackslash}p{0.135\textwidth}Y@{}}
\toprule
Category & Design & Scope & Core Mechanism & Impl. & Reported Metrics \\
\midrule
\multirow{3}{=}[-11.5pt]{Pipeline ext. \& iGPU} & GauRast~\cite{ref40} & Blending & Fixed-function eval. + blend & 28~nm; RTL+Layout & A: n.a.; S: n.a.\newline \mbox{24~FPS}; \mbox{1.7~W} \\
 & GBU~\cite{ref41} & Blending & Row cache + local blend & 28~nm; RTL+Sim & A: \mbox{0.90~mm$^2$}; S: \mbox{63~KB}\newline \mbox{92~FPS}; \mbox{0.22~W} \\
 & Vorion~\cite{ref43} & Blending, Bwd & Local color/opacity grad. accum. & 16~nm; Si+FPGA & A: \mbox{1.60~mm$^2$}; S: n.a.\newline \mbox{6.4--19~FPS}; \mbox{$<0.6$~W} \\
\midrule
\multirow{9}{=}[-36pt]{Dedicated accel.} & GSCore~\cite{ref44} & Forward & Subtile mask + hier. Sorting & 28~nm; RTL+Sim & A: \mbox{3.95~mm$^2$}; S: \mbox{272~KB}\newline \mbox{91.2~FPS}; \mbox{0.87~W} \\
 & GS-TG~\cite{ref45} & Forward & Cross-tile depth-order reuse & 28~nm; RTL+Sim & A: \mbox{3.984~mm$^2$}; S: \mbox{336~KB}\newline Relative; \mbox{1.063~W} \\
 & FLICKER~\cite{ref46} & Forward & Hierarchical subtile contribution tests & 28~nm; RTL+Sim & A: \mbox{3.47~mm$^2$}; S: \mbox{288~KB}\newline Relative \\
 & 129FPS Full-HD~\cite{ref47} & Forward & Low-SH pixel filtering & 28~nm; Gate & A: \mbox{0.66~mm$^2$}; S: \mbox{120~KB}\newline \mbox{129~FPS}; \mbox{0.219~W} \\
 & \mbox{STREAMINGGS}~\cite{ref70} & Forward & Voxel filter before attribute fetch & 32~nm; RTL+Sim & A: \mbox{5.37~mm$^2$}; S: \mbox{355~KB}\newline Relative \\
 & Neo~\cite{ref48} & Sorting & Incremental prior-frame order repair & 7~nm; RTL+Sim & A: \mbox{0.387~mm$^2$}; S: \mbox{264~KB}$^{\dagger}$\newline \mbox{99.3~FPS}; \mbox{0.798~W} \\
 & GCC~\cite{ref71} & Forward & Early rejection before attribute fetch & 28~nm; RTL+Sim & A: \mbox{2.711~mm$^2$}; S: \mbox{190~KB}\newline \mbox{667~FPS}; \mbox{0.79~W} \\
 & 3DGauCIM~\cite{ref75} & Forward & Digital CIM + near-mem. accum. & 16~nm; APR+Macro & A: \mbox{1.81/4.13~mm$^2$}; S: n.a.\newline \mbox{211~FPS}; \mbox{0.28/0.63~W} \\
 & REACT3D~\cite{ref49} & Fwd, Bwd & Block-selective updates + Fwd fallback & 12~nm; RTL+Sim & A: \mbox{12.45~mm$^2$}; S: \mbox{896~KB}$^{\dagger}$\newline \mbox{29.15~FPS}; \mbox{3.55~W} \\
\bottomrule
\end{tabularx}
\par\smallskip
\parbox{\textwidth}{\footnotesize Fwd = Projection, Sorting, Blending. Bwd = Backward. A = area. S = SRAM. Gate = gate-level synthesis. APR+Macro = place-and-route with macro modeling. The second line gives throughput and power. $^{\dagger}$ marks summed buffer components, including CAM in REACT3D. Relative indicates that no absolute result was reported. Metrics are retained from original papers across distinct nodes, scenes, and resolutions and are not directly comparable across rows. GBU uses different settings for quality and throughput. 3DGauCIM lists static and dynamic area and power and reports throughput for dynamic scenes.}
\end{table*}

Fig.~\ref{fig:runtime-workload-execution-synthesis}(d) illustrates where a single workload counter tracks runtime speedup closely and where it diverges under matched within-source conditions. FlashGS~\cite{ref38} and HiGS~\cite{ref55} achieve large speedups alongside sharp reductions in associations and list traversal, whereas Local-GS~\cite{ref66} leaves association counts nearly unchanged yet accelerates rendering through warp-level culling, scheduling, and parameter reuse. Faster-GS~\cite{ref96}, gsplat~\cite{ref37}, and CLM~\cite{ref61} illustrate a complementary boundary: when an optimization trims a single workload count or memory object while timing spans the complete Forward pass or full training iteration, the observed end-to-end speedup falls below the local work-reduction ratio. Evaluating GPU runtimes therefore requires pairing workload counters with identical timing boundaries so that work removal, execution reorganization, and data-movement costs are attributed accurately.

\section{Architectural Support for 3DGS}
\label{sec:arch}
\label{sec:architecture}

Even after runtime scheduling and list compaction are tuned, general-purpose GPUs remain constrained by irregular depth sorting, scattered attribute fetches, divergent pixel traversal, and atomic gradient contention. Hardware studies respond by extending existing graphics and compute pipelines or by designing dedicated architectures for Projection, Sorting, Blending, and Backward. Because reported area, power, and throughput figures span different technology nodes and often cover only a subset of stages, comparing hardware designs requires examining which operations move into specialized units, which intermediate lists stay on chip, and what work remains on programmable cores (Table~\ref{tab:accelerators}).

\subsection{Projection and Sorting Support}
\label{sec:arch_sort}

Projection determines the screen-space footprint of each active Gaussian, and Sorting expands those footprints into per-tile lists ordered by depth. On general-purpose hardware, generating, buffering, and sorting variable-length association lists consumes substantial DRAM bandwidth. Architectural designs reduce this pressure either by shrinking candidate lists before sorting or by reusing ordering across tiles and frames. GSCore~\cite{ref44} combines precise shape intersection, hierarchical sorting, and subtile masks to eliminate uncovered regions early. GS-TG~\cite{ref45} amortizes sorting across neighboring tiles through a shared depth order while rasterizing each tile independently. STREAMINGGS~\cite{ref70} applies voxel ordering and hierarchical filtering before fetching complete attributes, cutting both list volume and unnecessary memory reads.

Temporal coherence offers a complementary way to avoid sorting every frame from scratch. Neo~\cite{ref48} incrementally repairs the previous frame's depth lists as the camera moves, and GSAcc~\cite{ref86} exploits historical depth order to overlap pipeline stages. LS-Gaussian~\cite{ref72} reuses unchanged image regions across consecutive views and schedules the remaining tiles according to predicted workload. Addressing pipeline hazards inside execution itself, DeGS~\cite{ref85} decouples workload parsing and repacks dependent tasks into conflict-free issue packets.

\subsection{Blending and Backward Support}
\label{sec:arch_blend}
\label{sec:blending-backward-support}

Once depth-ordered lists are available, Blending evaluates candidate responses and accumulates pixel color and transmittance. Dedicated blending engines keep active pixel accumulators local while exploiting attribute reuse across pixels. GauRast~\cite{ref40} augments a GPU triangle rasterizer with fixed-function Gaussian evaluation and accumulation, leaving preprocessing and Sorting on CUDA cores. GBU~\cite{ref41} pairs a row-organized tile engine with a Gaussian reuse cache and local pixel buffers to accelerate ordered candidate processing at edge power budgets. Extending GSCore to binocular rendering, Nebula~\cite{ref99} (Section~\ref{sec:runtime_placement}) shares preprocessing and Sorting between both eyes while adding stereo reprojection, dedicated buffers, and list merging.

Several designs map Gaussian evaluation onto existing graphics pipelines or matrix engines to avoid standalone accelerators. RayGS~\cite{ref68} expresses splats via support quads executed on standard shaders and fixed-function alpha blending. DHR~\cite{ref67} maintains partial color and transmittance through programmable blending hardware and supports part of Backward. VR-Pipe~\cite{ref94} adds hardware early termination, multigranular tile binning, and quad merging to curb fragment shading overhead. Moving to neural accelerators, ORANGE~\cite{ref42} reformulates Gaussian-pixel evaluation as matrix operations on NPU arrays and uses tile batching to balance execution.

Hardware support for training faces an asymmetric dataflow challenge in Backward, where gradients from many pixels converge onto shared Gaussians. Vorion~\cite{ref43} integrates a Gaussian rasterizer into a RISC-V GPU to accumulate color and opacity gradients locally while leaving preprocessing, Sorting, and remaining updates on SIMT cores. Cambricon-GS~\cite{ref87} decouples Gaussian-wise and pixel-wise computation to balance hybrid parallelism during training. GauSPU~\cite{ref88}, GSArch~\cite{ref73}, and BOA-3DGS~\cite{ref89} employ local gradient merging, backward striding, or dedicated accumulation buffers to suppress irregular off-chip gradient writes. Architecturally, pixel-stationary buffers that hold color and transmittance during $N_{\text{eval}}$ and $N_{\text{blend}}$ serve a different purpose from the storage and reduction routing needed to combine gradients for the same Gaussian.

\subsection{Conditional Execution for Sparse Workloads}
\label{sec:arch_sparse}
\label{sec:conditional-execution}

Conditional execution uses lightweight early tests to skip uncovered regions or negligible contributions before expensive memory fetches or arithmetic begin. FLICKER~\cite{ref46} applies hierarchical contribution tests over smaller tiles to discard low-impact pixel work. The 129FPS Full-HD accelerator~\cite{ref47} combines hardware pixel filtering with pre-pruned, quantized, low-SH representations to meet full-HD throughput within a compact silicon budget.

Along the Forward path, the exact point of rejection determines which hardware cost is actually avoided. Gaussian-level filtering lowers $N_{\text{active}}$, shrinking downstream $N_{\text{pair}}$ and attribute traffic. Gaussian-tile tests reduce $N_{\text{pair}}$ directly and relieve Sorting storage and bandwidth. Pixel-level tests lower $N_{\text{eval}}$ only when applied before response evaluation. Once evaluation has started, the test already counts toward $N_{\text{eval}}$, and rejecting a candidate lowers $N_{\text{blend}}$ relative to a baseline only when the test discards contributions that the baseline would otherwise composite.

Across consecutive frames, online reconstruction and SLAM systems exhibit additional dynamic sparsity. REACT3D~\cite{ref49} runs a full Forward pass on each new keyframe, restricts subsequent updates to selected pixel blocks, and falls back to a full Forward pass whenever prediction error grows. Splatonic~\cite{ref78}, RTGS~\cite{ref22} (Section~\ref{sec:dynamic}), and Garnet~\cite{ref92} leverage sparse sampling, keyframe gating, gradient similarity, or historical state to curb redundant tracking and mapping operations. Upstream of rendering itself, SCARF~\cite{ref102} probes local geometry in its depth-guided encoder to skip redundant Gaussian generation.

\subsection{On-Chip Data and Cross-Frame Reuse}
\label{sec:arch_reuse}

Holding attributes and intermediate lists on chip eliminates repeated round-trips to external DRAM across stages. GCC~\cite{ref71} streams attributes through the pipeline and suppresses downstream reads immediately after a failed contribution test. STREAMINGGS~\cite{ref70} filters candidates using compact geometry and voxel metadata before fetching full appearance attributes for accepted Gaussians. Neural-3DGS Processor~\cite{ref90} reuses voxel parameters and features across neighboring tiles. The first two designs avoid fetching data that later stages will discard, whereas cross-tile reuse avoids fetching the same active attributes repeatedly.

Other architectures move computation closer to memory or organize external access around Gaussian locality. GSPIM~\cite{ref101} filters and compacts active 4D Gaussian records near memory before the GPU retrieves their full attributes. 3DGauCIM~\cite{ref75} maps Gaussian response evaluation onto digital compute-in-memory arrays and accumulates transmittance in near-memory units. IRIS~\cite{ref74} structures on-chip buffers and external memory transactions around Gaussian spatial locality so repeatedly referenced data stay near the processing elements.

Reuse can also extend across consecutive frames to bypass redundant sorting and shading. Neo~\cite{ref48} updates prior-frame depth lists incrementally under viewpoint motion. Lumina~\cite{ref91} caches both predicted depth order and rendered radiance: its radiance cache indexes pixel values using the identifiers of the first few Gaussians along a ray whose contributions exceed a threshold, reusing the cached pixel on a hit and falling back to rasterization on a miss. In SLAM pipelines, DF-SAR~\cite{ref93} preserves alpha values, voxel data, or keyframe states across frames to trim repeated rendering and mapping traffic.

\begin{figure*}[!tp]
  \centering
  \includegraphics[width=0.96\textwidth]{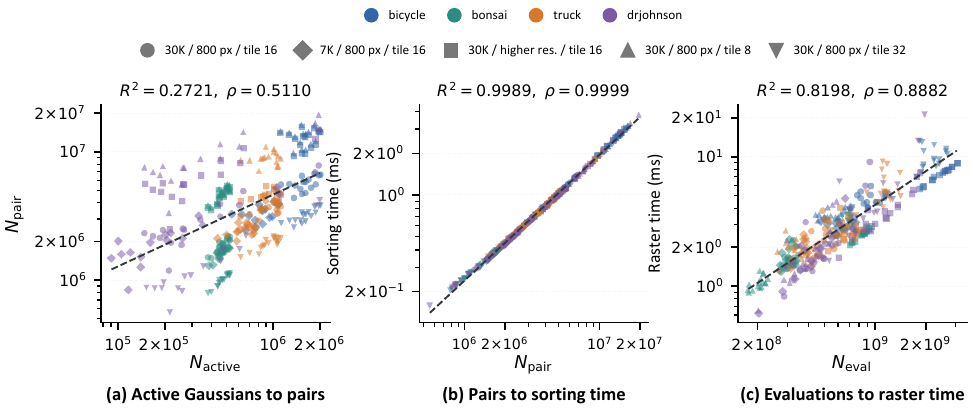}
  \caption{Workload and kernel-timing relationships under controlled conditions. The panels connect active Gaussians to associations, associations to Sorting time, and pixel evaluations to rasterization time within Blending. Each panel uses 16 views of each of four scenes under five configurations, for 320 configuration--view measurements. The higher-resolution setting is $1600\times1063$ for Bicycle, $1600\times1066$ for Bonsai, $979\times546$ for Truck, and $1332\times876$ for DrJohnson. $R^2$ is the coefficient of determination for an ordinary least-squares fit of the base-10 log response on the base-10 log predictor, and $\rho$ is Spearman rank correlation.}
  \label{fig:unified-workload}
\end{figure*}

\subsection{Architectural Design Space}
\label{sec:arch_space}

The central trade-off in 3DGS hardware design lies between energy efficiency and extensibility. Across their reported evaluation scopes in Table~\ref{tab:accelerators}, dedicated Forward accelerators and blending plug-ins operate within 0.2--1.7~W and under 5.4~mm$^2$, far below full desktop GPU power envelopes. These savings stem from keeping intermediate lists and active attributes on chip and pruning redundant evaluations in fixed-function datapaths, yet specialization introduces substantial design complexity and rigid assumptions. Supporting irregular, variable-length lists alongside training traffic and online SLAM states further expands control logic and on-chip storage, as illustrated by REACT3D's 12.45~mm$^2$ footprint and 896~KB of on-chip buffering (including 64~KB of CAM). At the same time, hardening fixed dataflows locks in specific Gaussian parameterizations, SH degrees, and alpha-compositing rules. As algorithmic research advances toward compressed codebooks, neural appearance fields, dynamic deformations, and relaxed transparency semantics (Section~\ref{sec:algo}), overly rigid accelerators become difficult to extend and risk frequent fallbacks to a host processor.

Striking a practical balance between generality and specialization depends on pairing deployment power budgets with the stability of the target workload. On tightly power-constrained edge and XR devices dominated by standard Forward rendering, dedicated Forward or near-memory architectures such as GSCore, GCC, STREAMINGGS, 129FPS Full-HD, 3DGauCIM, and IRIS are well suited because the projection-sorting-blending core is relatively stable and cutting DRAM round-trips is necessary to operate near a sub-watt envelope. On desktop GPUs, mobile SoCs, and multi-purpose platforms that must support rapidly evolving 3DGS variants alongside conventional triangle rendering, lightweight pipeline extensions and existing-accelerator mappings such as GauRast, GBU, VR-Pipe, and ORANGE offer a stronger trade-off: they accelerate the stable, compute-dense inner loop of Gaussian evaluation and accumulation at low silicon cost while leaving visibility culling, attribute decoding, deformation, and compositing policies programmable. For on-device training and dense SLAM, where backward gradient reduction, optimizer updates, and dynamic map management carry heavy buffering demands and continue to evolve rapidly, hybrid architectures such as Vorion, GSArch, Cambricon-GS, and REACT3D define a more resilient boundary by specializing contention-heavy gradient accumulation and sparse filtering while keeping optimizer states and update control on programmable cores. Across these deployment settings, end-to-end gains depend on how specialized units change memory traffic and interact with the remaining stages. Section~\ref{sec:findings} uses matched GPU measurements to examine how workload and data movement affect complete Forward latency and DRAM traffic.

\section{Workload Findings and Implications}
\label{sec:findings}
\label{sec:synthesis}

Sections~\ref{sec:algo}--\ref{sec:arch} examine how algorithms, GPU runtimes, and silicon architectures alter the 3DGS execution path, yet isolated metrics such as compression ratio, Gaussian count, or local kernel speedup rarely predict end-to-end latency, memory footprint, and tail stability. Combining this full-stack taxonomy with controlled GPU profiling via GSPROF yields three cross-layer findings that connect how attributes are stored and moved, how spatial work is grouped, and which operations limit performance once rasterization speeds up.

\subsection{Workload Reduction Must Reach the Backend}
\label{sec:findings_backend}
\label{sec:materialization}

Reducing stored model size or visible Gaussian counts improves end-to-end speed only when those savings reach actual backend computation and memory transfers without being offset by extra data copying or decoding. To understand where upstream counts succeed or fail as predictors, Fig.~\ref{fig:unified-workload} measures four scenes across training checkpoints, output resolutions, and tile sizes, linking active Gaussians, Gaussian--tile pairs, pixel evaluations, and stage times within a single codebase. Across this sweep, the number of active Gaussians correlates only moderately with the number of tile pairs because projected footprints vary with camera distance and scene geometry. Once pairs and pixel evaluations are formed, Sorting time tracks $N_{\text{pair}}$ closely, as does rasterization time with $N_{\text{eval}}$.

\begin{figure*}[t]
  \centering
  \includegraphics[width=0.96\textwidth]{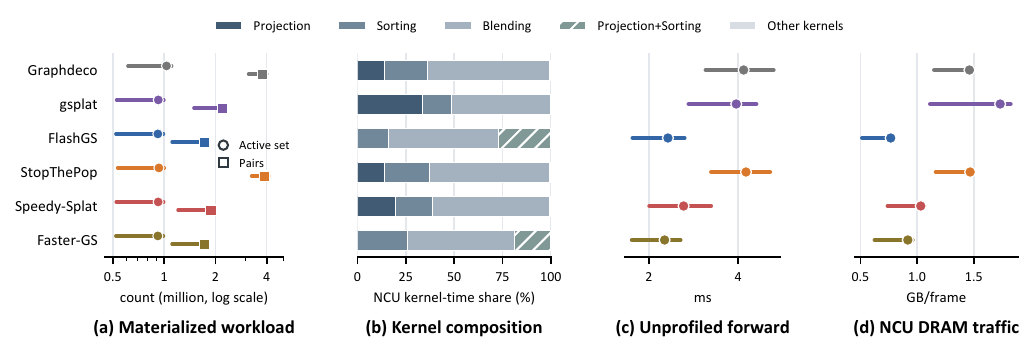}
  \caption{Six 3DGS backends on the same four Tanks \& Temples Truck views at $800\times446$, with a 30K-iteration model containing 2,541,226 resident Gaussians. Panel (a) shows active Gaussians and materialized pairs; panel (b) groups kernels by Projection, Sorting, and Blending, keeping cross-stage fused kernels joint. Panels (c)--(d) show independently measured complete Forward latency and DRAM traffic. Markers are medians over views, and lines span minima to maxima.}
  \label{fig:workload-objects}
\end{figure*}

Even strong downstream counts can mislead when different renderers prepare scene data in different ways. Comparing six backends on identical Truck views in Fig.~\ref{fig:workload-objects} shows that gsplat generates roughly 42\% fewer tile pairs than Graphdeco at the median, yet complete Forward latency drops by only 4\% while total DRAM traffic rises by 19\%. Fig.~\ref{fig:cross-scene-backend-quality}(a) extends the Graphdeco-versus-gsplat comparison to 127 test photographs across four scenes at width 800. On every view, gsplat emits fewer tile pairs than Graphdeco, yet it renders 28 views more slowly, most notably in Bonsai, where the majority of views slow down despite far fewer pairs. Hardware-counter profiles show higher total DRAM traffic for gsplat on all 12 profiled Bicycle, Bonsai, and Truck views and lower traffic on all four profiled DrJohnson views. \supsec{3} reports the full six-backend results across the 127 views.

One source of this discrepancy is how each renderer supplies appearance attributes for color shading. Graphdeco evaluates color directly inside Projection, whereas standard packed gsplat copies visible positions, opacities, and SH coefficients into temporary arrays before shading, generating roughly a quarter of its total memory traffic. Table~\ref{tab:sh3} tests whether eliminating this temporary copy recovers speed and bandwidth under matched rendering settings. It compares three full-SH3 gsplat layouts: packed execution with temporary copies, dense execution via \texttt{packed=False} that reads directly from full resident arrays, and indexed shading that fetches resident SH coefficients on demand by Gaussian index. Across all eight configurations, indexed shading lowers Forward P95 latency by 8.2--14.3\% over packed execution and achieves the lowest peak memory allocation. Dense execution marginally outperforms indexed shading at P95 in three scene and resolution settings, yet requires up to about five times as much temporary memory on Bicycle.

Expanding the matched copy ablation to 40 photo-anchor and resolution conditions shows that indexed shading outperforms packed execution in every case. As Fig.~\ref{fig:cross-scene-backend-quality}(b) shows, eliminating the temporary SH array accounts for a median of 74.8\% of the latency gain, and indexed shading cuts physical DRAM traffic relative to the matched copy variant with fused shading by 6.7--23.7\%, with a median of 17.9\%. The packed-to-indexed time saving is larger for SH3 than for SH0 at all 40 conditions. This comparison changes both attribute size and shading arithmetic, as detailed in \supsec{4.1}.

Algorithmic compression exhibits the same trade-off over the lifetime of a scene. On the released CompGS Truck models~\cite{compgsVQ}, compact GPU codebooks expand more than sevenfold once decoded. Across all 251 views at two output resolutions, decoding coefficients every frame rather than once at load time adds roughly 3.4~ms per frame for identical output pixels, nearly doubling median frame time at width 800, as documented in \supsec{4.2}. The desktop backends in Fig.~\ref{fig:workload-objects} and the hardware designs in Table~\ref{tab:accelerators} illustrate a shared principle across all three layers. Screen compact geometry first and decode or fetch full appearance attributes only for surviving Gaussians. This avoids materializing short-lived, wide attributes. For compression in Section~\ref{sec:compaction}, the benefit depends on when decoded attributes are needed. In GPU runtimes (Section~\ref{sec:runtime}), reading resident attributes by index can avoid a temporary copy. GCC~\cite{ref71} and STREAMINGGS~\cite{ref70} apply the same principle in hardware (Section~\ref{sec:arch_reuse}).

\subsection{Optimal Granularity Is Stage Dependent}
\label{sec:findings_granularity}

Uniform tiling trades Sorting work against Blending work, with the lowest median Forward latencies at 12 or 16 pixels in our sweep. The effect of a fixed screen-space radius threshold depends on scene detail, resolution, and camera trajectory. Fig.~\ref{fig:runtime-workload-execution-synthesis}(a)--(c) sweeps tile widths from 8 to 32 pixels on identical views across four scenes at width 800. Small tiles cut pixel evaluations and rasterization time but increase tile pairs and Sorting overhead. Large tiles reverse that balance while leaving blended contributions nearly unchanged. The two-width sweep in \supsec{4.3} finds one of these intermediate optima in all eight scene and resolution combinations. An intermediate width outperforms both extremes on nearly every tested view. Because a uniform grid forces Sorting and Blending into a compromise, HiGS~\cite{ref55} sorts on coarse macro tiles before rasterizing fine tiles, GS-TG~\cite{ref45} shares depth order across tiles, and GSCore~\cite{ref44} and FLICKER~\cite{ref46} filter at subtile granularity. Each design matches a stage to the scale at which it can share work or reject contributions.

\begin{table*}[tp]
\caption{Full-SH3 Layouts and the Cost of Materializing Coefficients}
\label{tab:sh3}
\centering\scriptsize\setlength{\tabcolsep}{3pt}
\begin{tabular*}{\textwidth}{@{\extracolsep{\fill}}lrrrrrrrrrrrr@{}}
\toprule
 & & \multicolumn{3}{c}{Forward P95 (ms)} & \multicolumn{3}{c}{Peak allocation (GiB)} & \multicolumn{2}{c}{Gather (ms)} & \multicolumn{2}{c}{SH kernel (ms)} & SH array \\
Scene & Width & Dense & Packed & Indexed & Dense & Packed & Indexed & Packed & Indexed & Packed & Indexed & share (\%) \\
\midrule
Bicycle & 800 & 5.941 & 6.312 & 5.412 & 1.837 & 1.942 & 1.597 & 1.271 & 0.068 & 0.458 & 0.511 & 76.6 \\
Bicycle & 1920 & 8.091 & 8.539 & 7.606 & 1.905 & 2.003 & 1.658 & 1.237 & 0.068 & 0.458 & 0.510 & 77.3 \\
Bonsai & 800 & 2.661 & 2.933 & 2.658 & 0.393 & 0.451 & 0.357 & 0.305 & 0.019 & 0.109 & 0.121 & 67.5 \\
Bonsai & 1920 & 3.423 & 3.774 & 3.465 & 0.434 & 0.497 & 0.403 & 0.304 & 0.019 & 0.109 & 0.121 & 71.9 \\
Truck & 800 & 3.348 & 3.724 & 3.201 & 0.783 & 0.895 & 0.705 & 0.707 & 0.040 & 0.255 & 0.284 & 75.2 \\
Truck & 1920 & 4.749 & 5.185 & 4.608 & 0.889 & 0.943 & 0.769 & 0.705 & 0.040 & 0.255 & 0.283 & 74.3 \\
DrJohnson & 800 & 7.969 & 8.935 & 8.034 & 1.071 & 1.317 & 0.993 & 0.187 & 0.016 & 0.062 & 0.072 & 51.6 \\
DrJohnson & 1920 & 7.968 & 9.139 & 7.997 & 1.186 & 1.410 & 1.104 & 0.187 & 0.016 & 0.062 & 0.072 & 48.4 \\
\bottomrule\end{tabular*}
\par\smallskip\raggedright This layout and memory suite uses one 97-pose route per scene. P95 summarizes per-pose medians. Peak allocation includes resident tensors and is the maximum over all 97 poses. Gather and SH times are medians of five anchor-zero replays. SH array share is $(t_M-t_I)/(t_P-t_I)$ from 30 unprofiled Forward calls per variant at that anchor. Here $t_P$ is the standard packed time, $t_I$ is the indexed time, and $t_M$ is the time for an intermediate variant that materializes the packed SH array and uses the same fused shading kernel as indexed execution. \supsec{4.1} reports the separate 40-condition DRAM verification. Section~\ref{sec:findings_granularity} and \supsec{4.3} use a separate 12-route timing suite with 30 calls per mode and pose.
\end{table*}

\begin{figure*}[t]
  \centering
  \includegraphics[width=\textwidth]{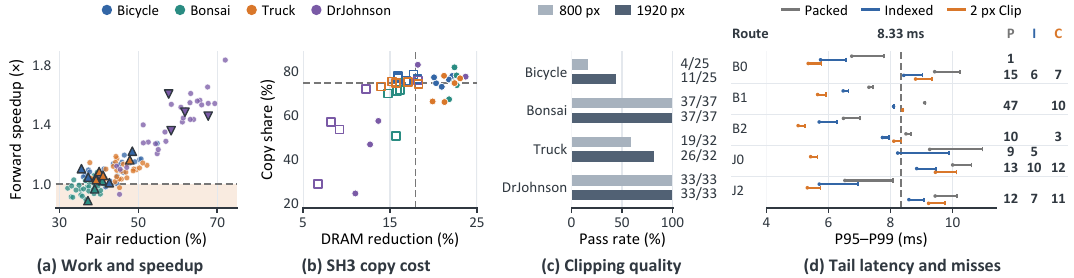}
  \caption{Matched Forward measurements. (a) Graphdeco versus gsplat at width 800. Up/down triangles indicate higher/lower DRAM traffic for gsplat. (b) DRAM reduction and share of latency savings from removing the SH3 copy. Dashed lines mark medians. Circles and squares denote widths 800 and 1920, respectively. (c) Quality pass rates with 2-pixel clipping. (d) Cross-pose P95--P99 latency and 120~Hz deadline misses, with width 800 above 1920 for each route.}
  \label{fig:cross-scene-backend-quality}
\end{figure*}

Fixed screen-space culling rules exhibit an analogous limitation under joint visual-quality and resolution constraints. Fig.~\ref{fig:cross-scene-backend-quality}(c) evaluates 2-pixel clipping against PSNR, SSIM, and LPIPS bounds relative to packed execution. Clipping Gaussians with projected radii up to 2 pixels meets all three quality criteria on only 93 of 127 test photographs at width 800 and 107 at width 1920, with failures on Bicycle and Truck. Indexed and dense execution meet the bounds on every photograph at both widths. Cross-pose P95 and P99 latencies along 12 camera trajectories at both widths show why single-view averages obscure missed deadlines under a 120~Hz budget. Indexed shading lowers Forward P95 latency relative to packed execution across all 24 route and width combinations, while the missed-frame counts in Fig.~\ref{fig:cross-scene-backend-quality}(d) depend on both resolution and camera path. On the original DrJohnson trajectory at width 800, 2-pixel clipping eliminates missed frames, while indexed shading meets the budget at P95 but exceeds it at P99. At width 1920, indexed shading misses fewer frames than clipping on that trajectory. Four other routes, including all three Bicycle trajectories, tie at width 800 and favor indexed at width 1920. The remaining seven tie at both widths. Stage dominance also varies by view. On one high-resolution Truck frame, Graphdeco spends more time in radix Sorting than in rasterization. \supsec{4.3} reports the quality, trajectory, and stage-breakdown results.

Across algorithmic culling in Sections~\ref{sec:activation}--\ref{sec:screenspace}, GPU tiling in Section~\ref{sec:runtime}, and subtile hardware in Section~\ref{sec:arch}, spatial grouping and filtering cannot be tuned with a single static parameter. Effective systems must match granularity to each stage and resolution while verifying both visual fidelity and worst-case frame deadlines.

\subsection{Local Speedups Shift the Bottleneck}
\label{sec:findings_bottleneck}

Once Forward rasterization becomes fast, parameter updates, memory transfers, and cross-frame recovery can limit end-to-end performance. This shift affects workloads beyond static single-frame inference across all three layers. In training, once Faster-GS~\cite{ref96} curtails rasterization overhead, Adam updates absorb a substantial share of each iteration. Fusing updates into Backward saves time and memory, yet Update work continues to limit scaling. In the dynamic scenes and online SLAM settings described in Section~\ref{sec:dynamic}, SpeeDe3DGS~\cite{ref20} and RTGS~\cite{ref22} address analogous recurring costs in per-frame deformation and map updates. Because an increasing stage share can simply reflect faster neighboring stages, reporting absolute stage times alongside complete-iteration latency distinguishes genuine slowdowns from shifted bottlenecks.

Data movement and temporal reuse expose the same need to account for the complete execution path. When scenes outgrow GPU memory or stream over a network, TideGS~\cite{ref98} overlaps SSD, CPU, and GPU transfers while Nebula~\cite{ref99} streams newly visible Gaussians from cloud to client, tying net speed to sustained transfer efficiency across complete sessions. Hardware accelerators that reuse prior frames in Section~\ref{sec:arch_reuse} incur comparable recovery costs. Neo~\cite{ref48} repairs outdated depth lists, Lumina~\cite{ref91} falls back to rasterization when its radiance cache misses, and REACT3D~\cite{ref49} repeats full Forward passes whenever prediction drifts. Across these designs, net speedup depends on whether avoided work outweighs the cost of updates, transfers, cache maintenance, and fallback passes.

Taken together, these three findings establish practical rules for evaluating 3DGS systems. Layout and compression changes should hold geometry and rendered output fixed while measuring complete Forward latency, peak memory, and physical DRAM traffic. Tiling and culling policies should pair explicit visual-quality bounds with P99 latency and missed-deadline counts across multiple trajectories at each target resolution. Training, offloading, and temporal-reuse methods should include updates, transfers, and fallback paths inside the timed window. Cache size, working-set locality, and memory bandwidth can shift the crossover between copying attributes and reading them by index. In our controlled profiling, both variants in each paired comparison run on the same NVIDIA RTX~3090 for monocular Forward rendering, excluding compositor and display overhead. \supsec{2} documents GPU assignments and replay protocols.

\section{Evaluation Principles and Open Issues}
\label{sec:eval_open}
\label{sec:evaluation}

The cross-layer workload effects in Section~\ref{sec:findings} explain why isolated benchmarks routinely mislead system design. Establishing comparable efficiency claims requires aligning visual-quality constraints, workload traceability, and sustained execution windows, while also clarifying the structural problems that remain open for future 3DGS systems.

\subsection{Evaluation under Quality Constraints}
\label{sec:eval_quality}

Efficiency comparisons are meaningful only when tied to verified visual output. Compression and pruning introduce reconstruction error, StopThePop~\cite{ref27} addresses multi-view popping consistency, and WSR~\cite{ref59} and StochasticSplats~\cite{ref60} alter transparency semantics. Training introduces an additional convergence constraint: Faster-GS~\cite{ref96} treats skipping updates for invisible Gaussians as optional because deviating from strict Adam updates can shift final reconstruction quality. Matched hyperparameters therefore do not guarantee matched fidelity, and faster per-iteration timing does not establish equivalent convergence.

Comparison claims must also stay within the exact scope supported by each evidence set. Figs.~\ref{fig:compression-speed} and~\ref{fig:runtime-workload-execution-synthesis}(d) pair baseline and optimized configurations strictly within each source, tying storage or workload reductions to speedup at the corresponding quality point. Fig.~\ref{fig:workload-objects} and \supsec{3} verify per-view image fidelity under a fixed model and view set to support backend comparisons within that scope. Neither evidence class justifies ranking separate studies evaluated under different quality budgets.

Even when camera matrices and scene checkpoints are identical, undocumented rendering defaults can confound cross-backend comparisons. On an interpolated camera pose at width 800, aligning gsplat's near clipping plane from 0.01 to Graphdeco's default of 0.2 reduces the mean absolute RGB difference between their rendered images from 0.169 to $1.3 \times 10^{-4}$ on Truck and from 0.108 to $3.4 \times 10^{-5}$ on Bicycle. Extending this check across 1,040 trajectory conditions spanning 12 scenes confirms that the same near-plane alignment reduces mean absolute RGB error by more than $10^{-5}$ on at least one pose in 508 conditions, with zero conditions degrading beyond that threshold, as documented in \supsec{4.4}. Verifying output parity before timing and recording clipping conventions alongside camera parameters, compositing semantics, and per-anchor quality thresholds prevents such defaults from silently skewing efficiency comparisons.

\begin{figure*}[!t]
  \centering
  \includegraphics[width=0.93\textwidth]{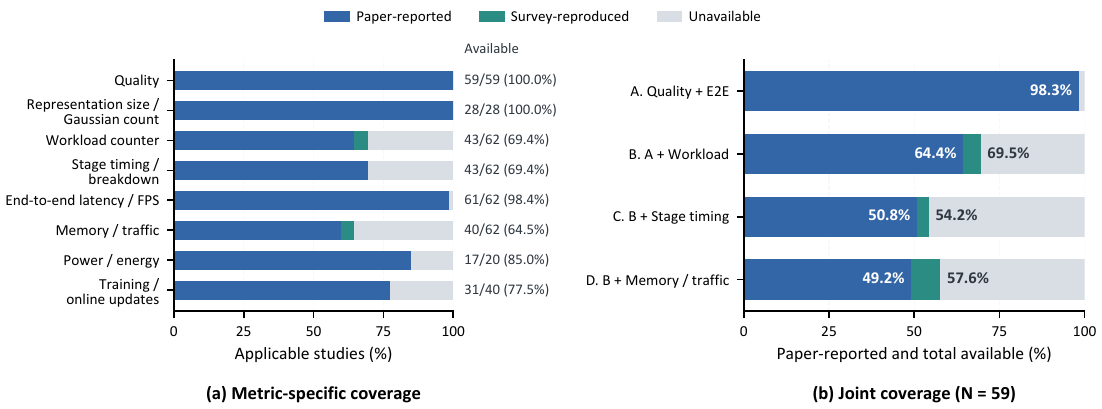}
  \caption{Evidence coverage in the audited literature subset. Panel (a) shows the available count over the number of applicable studies for each metric, with the corresponding percentage. Panel (b) uses a common denominator of $N=59$ studies with applicable quality comparisons. Both panels distinguish paper-reported evidence from additions reproduced by this survey and unavailable evidence.}
  \label{fig:coverage}
\end{figure*}

\subsection{Traceable Workloads and Sustained Operation}
\label{sec:eval_traceable}

Comparable timing also requires recording what generates the executed workload. Hierarchical and multiscale schemes~\cite{ref17,ref28,ref30} alter $N_{\text{active}}$, whereas spatial filtering in FlashGS~\cite{ref38} shrinks $N_{\text{pair}}$. While Fig.~\ref{fig:unified-workload} isolates work-to-time scaling under fixed code and Fig.~\ref{fig:workload-objects} isolates backend implementation differences under fixed inputs, our audit of 59 applicable efficiency studies in Fig.~\ref{fig:coverage} exposes a persistent reporting gap: 98.3\% jointly report visual quality and end-to-end performance, yet only 50.8\% originally pair those figures with workload counts and stage timing and only 49.2\% with workload and memory or traffic measurements. Even after incorporating evidence reproduced in this survey, joint availability reaches only 54.2\% and 57.6\%. A practical barrier to closing this gap is that attaching hardware profilers or counter instrumentation directly to latency runs perturbs measured execution time. Decoupling unprofiled complete-path timing from lightweight counter extraction and separate diagnostic profiler replays, as structured in GSPROF, allows quality, physical workload counts, stage timing, and memory traffic to be reported over aligned inputs without distorting latency.

Sustained operation adds temporal and thermal dimensions that single-frame averages obscure. Viewpoint trajectories govern depth-list repair in Neo~\cite{ref48} and full-Forward fallbacks in REACT3D~\cite{ref49}. Training view order shapes out-of-core working sets in TideGS~\cite{ref98}. Cloud-client interactivity bounds Nebula~\cite{ref99}. Mobile-GS~\cite{ref34} distinguishes cold runs from thermally constrained sustained performance. As Fig.~\ref{fig:cross-scene-backend-quality}(d) illustrates for DrJohnson in Section~\ref{sec:findings_granularity}, a backend can satisfy a 120~Hz budget at P95 while still missing deadlines at P99. System evaluations should therefore replace isolated view averages with continuous trajectories and sustained operating windows, reporting P99 tail latency, missed-deadline counts, sustained power, and bursts of reordering, recomputation, or data transfer alongside the camera path and complete timed interval.

\subsection{Open Problems for 3DGS Systems}
\label{sec:open_problems}

\textbf{Shared workload interfaces.} Figs.~\ref{fig:workload-objects} and~\ref{fig:cross-scene-backend-quality}(a)--(b) show that distinct backends materialize incompatible associations, attribute buffers, and memory accesses from the exact same scene input. Offline instrumentation layers such as GSPROF can extract comparable workload counts across existing renderers for post-hoc attribution, yet eliminating redundant work during live execution requires shared interfaces across algorithms, runtimes, and accelerators that define canonical representations for active-Gaussian selection, tile coverage, sorted lists, and attribute lifecycles. Key open questions concern which intermediate states should be standardized, which culling and compositing rules must remain backend-specific, and whether eliminating repeated construction outweighs interface conversion and invalidation costs.

\textbf{Adaptive execution under varying workloads.} The tile-granularity crossover in Fig.~\ref{fig:runtime-workload-execution-synthesis}, the configuration spread in Fig.~\ref{fig:unified-workload}, and the resolution- and trajectory-dependent culling trade-offs in Section~\ref{sec:findings_granularity} demonstrate that static policies fail across scenes, trajectories, and resolutions. Stage-decoupled tiling in HiGS~\cite{ref55} and dynamic assignment in Balanced 3DGS~\cite{ref54} offer complementary starting points. Future systems need lightweight runtime observation of association volume, traversal length, and tile imbalance to adapt grouping, culling thresholds, and hardware resources online while bounding monitoring, switching, and list-rebuilding overhead.

\textbf{Programmable specialization boundaries.} As Section~\ref{sec:arch} shows, specialized hardware delivers substantial power reductions on stable Forward stages, yet training, dynamic scenes, compressed representations, and evolving compositing rules rapidly inflate control complexity and on-chip buffering. Determining which core primitives to harden in silicon and which representation, culling, and update policies to leave programmable requires whole-task evaluation across static rendering, incremental SLAM, and on-device training. Future architectures must balance low-power efficiency against functional coverage and algorithmic extensibility, evaluating complete workloads beyond isolated kernel throughput.

\onecolumn
{\setlength{\multicolsep}{0pt}%
\begin{multicols}{2}
\section{Conclusion}
\label{sec:conclusion}

This survey unifies 3DGS algorithms, GPU runtimes, and architectural support on a shared execution path through a workload-centric framework. Because model size, Gaussian counts, and local kernel speedups capture isolated points along that path, no single metric predicts end-to-end efficiency. Combining full-stack synthesis with controlled GPU profiling via GSPROF shows that real system gains depend on three coupled conditions: upstream reductions must reach backend execution without being offset by attribute decoding or intermediate materialization, execution granularity and culling policies must adapt across stages, resolutions, and trajectories under strict visual-quality and tail-latency bounds, and local rasterization speedups must be weighed against shifted bottlenecks in parameter updates, tiered transfers, and cross-frame reuse.

This workload-centric perspective grounds practical 3DGS evaluation in explicit quality constraints, traceable physical counts, aligned timing boundaries, and sustained trajectory measurements. It also frames three enduring system challenges in shared cross-layer interfaces, workload-adaptive execution, and programmable hardware specialization. Tracing work continuously from explicit scene representations to silicon provides a rigorous foundation for both fair benchmarking and integrated 3DGS system design.

\end{multicols}}

\begingroup\small
\bibliographystyle{IEEEtran}
\bibliography{references}
\endgroup

\appendix
\section*{Supplemental Material}
\addcontentsline{toc}{section}{Supplemental Material}
\renewcommand{\thesection}{S\arabic{section}}
\setcounter{table}{0}
\renewcommand{\thetable}{S\arabic{table}}
\section{Survey Corpus Construction and Screening Criteria}
\label{app:survey-scope}
\label{app:arxiv-S1}

The survey covers publicly available research on efficient 3D Gaussian splatting through November 2026. The retained corpus draws primary records from IEEE Xplore, the ACM Digital Library, CVF Open Access, OpenReview, arXiv, and official proceedings, journal, supplementary, and project pages. Covered venue families include CVPR, ICCV, ECCV, SIGGRAPH, TOG, NeurIPS, ICLR, ASPLOS, HPCA, MICRO, DAC, DATE, and ICCAD. Search concepts combine ``3D Gaussian splatting'' or ``Gaussian splatting'' with terms for compression, pruning, level of detail, culling, rasterization, sorting, GPU execution, memory or communication, accelerators, training, dynamic scenes, SLAM, and mobile or edge systems.

Candidates were screened for a direct connection to 3DGS system efficiency. Algorithmic works are included only when they materially affect stored Gaussian data, resident or per-frame active Gaussians, per-frame rendering work, data movement, cached depth orders or image reuse, gradients and optimizer arrays, or an explicitly measured system-efficiency outcome. GPU-runtime and architecture studies are included when they implement or evaluate a stage on the rendering, training, or online-update path. Pure application papers and studies concerned solely with visual quality, reconstruction accuracy, or application metrics without a relevant system cost are excluded. We also exclude adjacent non-3DGS methods, duplicate versions without additional system evidence, and records whose primary full text or official supplement cannot be verified. When multiple versions of a study exist, the most complete verifiable version is retained while preserving its reported measurement scope.

Screening examines titles and abstracts followed by full texts and official supplements, distinguishing narrative citation from quantitative admission. Quantitative records must identify the affected workload or resource quantity, comparison baseline, input and configuration, quality settings, and timed operations. Missing fields are recorded as unavailable.

The main bibliography contains 104 entries, and Sections~3--5 together with their method tables cite 84 distinct works. Our reporting audit evaluates 62 efficiency studies, of which 59 include an applicable image-quality comparison and form the fixed denominator of main Fig.~\ref{fig:coverage}(b), while Fig.~\ref{fig:coverage}(a) uses the applicable denominator for each metric. Across these 59 studies, 58 provide visual quality and end-to-end performance at 98.3\%, 41 additionally provide workload counts at 69.5\%, 32 include stage timing at 54.2\%, and 34 include memory or traffic evidence at 57.6\%. Excluding additions reproduced by this survey, the original papers alone account for 58 of 59 studies at 98.3\%, 38 of 59 at 64.4\%, 30 of 59 at 50.8\%, and 29 of 59 at 49.2\%, respectively, with survey reproductions adding 0, 3, 2, and 5 records across these four joint conditions.

\subsection{Per-Paper Audit Checklist}
\label{app:reporting-checklist}

Table~\ref{tab:supp-audit-checklist} lists the 62 audited studies using method names aligned with the main article. Here, P denotes paper-reported evidence, R denotes evidence reproduced by this survey, U denotes applicable but unavailable evidence, and N/A denotes an inapplicable metric. The columns Rep., Work, Stage, E2E, Mem., and Train correspond to representation size or Gaussian count, workload counter, stage timing, end-to-end performance, memory or traffic, and training or online-state evidence, respectively. Quality and Power indicate image-quality and electrical-efficiency evidence. The checklist records the literature subset used in the reporting audit. Its membership is tracked independently of the article bibliography.

\par
\ifdim\dimexpr\pagegoal-\pagetotal\relax<9\baselineskip\newpage\fi
\begingroup\scriptsize
\setlength{\tabcolsep}{4pt}
\renewcommand{\arraystretch}{1.08}
\begin{longtable}{@{}r>{\raggedright\arraybackslash}p{0.47\textwidth}cccccccc@{}}
\caption{Per-Paper Evidence Checklist for Main Fig.~\ref{fig:coverage}}\label{tab:supp-audit-checklist}\\
\toprule
ID & Paper & Quality & Rep. & Work & Stage & E2E & Mem. & Power & Train \\
\midrule\endfirsthead
\toprule
ID & Paper & Quality & Rep. & Work & Stage & E2E & Mem. & Power & Train \\
\midrule\endhead
\bottomrule\endfoot
1 & Hierarchical 3D Gaussian Representation & P & P & P & P & P & P & N/A & P \\
2 & Reducing the Memory Footprint of 3D Gaussian Splatting & P & P & U & U & P & U & N/A & P \\
3 & MetaSapiens & P & P & P & P & P & U & P & P \\
4 & SpeeDe3DGS & P & P & U & U & P & U & N/A & P \\
5 & MEGS2 & P & P & U & U & P & P & N/A & U \\
6 & RTGS & P & P & P & P & P & P & P & P \\
7 & Monte Carlo Energy Aggregation (Flux-GS) & P & P & U & U & P & P & N/A & N/A \\
8 & Provable Pruning via Coresets & P & P & U & U & U & U & N/A & P \\
9 & Compressed 3D Gaussian Splatting & P & P & U & U & P & U & N/A & U \\
10 & Fast 3DGS Rendering via Easily Integrable Improvements & P & N/A & P & P & P & U & N/A & N/A \\
11 & StopThePop & P & P & R & P & P & R & N/A & P \\
12 & FLoD & P & P & P & U & P & P & N/A & U \\
13 & Compact 3D Gaussian Representation & P & P & U & U & P & U & N/A & U \\
14 & Multi-Scale 3D Gaussian Splatting & P & P & P & U & P & U & N/A & U \\
15 & LightGaussian & P & P & U & U & P & U & N/A & U \\
16 & HAC & P & P & U & P & P & U & N/A & P \\
17 & Trimming the Fat & P & P & U & U & P & U & N/A & U \\
18 & Mobile-GS & P & P & U & P & P & U & P & P \\
19 & High-Fidelity Mobile Avatars & P & P & P & U & P & P & N/A & P \\
20 & 3DGEER & P & N/A & P & P & P & U & N/A & U \\
21 & gsplat & P & N/A & R & U & P & P & N/A & P \\
22 & FlashGS & P & N/A & P & P & P & R & N/A & N/A \\
23 & Seele & P & N/A & P & U & P & P & N/A & N/A \\
24 & GauRast & N/A & N/A & U & P & P & U & P & N/A \\
25 & Gaussian Blending Unit & P & N/A & P & P & P & P & P & N/A \\
26 & Vorion & P & N/A & P & P & P & P & P & P \\
27 & GSCore & P & N/A & P & P & P & P & P & N/A \\
28 & GS-TG & N/A & N/A & P & P & P & P & P & N/A \\
29 & FLICKER & P & N/A & P & P & P & P & P & N/A \\
30 & 129FPS Full-HD 3DGS Accelerator & P & P & P & U & P & P & P & N/A \\
31 & Neo & P & N/A & P & P & P & P & P & N/A \\
32 & REACT3D & P & N/A & P & P & P & P & P & P \\
33 & HAC++ & P & P & U & P & P & P & N/A & P \\
34 & CodecGS & P & P & U & P & P & U & N/A & P \\
35 & TC-GS & P & P & U & P & P & U & N/A & P \\
36 & NeuralGS & P & P & U & P & P & U & N/A & P \\
37 & Balanced 3DGS & P & N/A & P & P & P & U & N/A & P \\
38 & HiGS & P & N/A & P & P & P & P & N/A & N/A \\
39 & ARC (DISTWAR) & N/A & N/A & P & P & P & U & N/A & P \\
40 & Taming 3DGS & P & P & P & P & P & P & N/A & P \\
41 & Hybrid Transparency & P & N/A & P & P & P & U & N/A & P \\
42 & Weighted Sum Rendering & P & N/A & U & P & P & P & N/A & N/A \\
43 & StochasticSplats & P & N/A & P & U & P & U & N/A & N/A \\
44 & CLM & P & N/A & P & P & P & P & N/A & P \\
45 & GS-Scale & P & N/A & P & P & P & P & N/A & P \\
46 & On Scaling Up 3DGS Training (Grendel) & P & N/A & P & P & P & P & N/A & P \\
47 & LiteGS & P & P & P & P & P & P & N/A & P \\
48 & LoBE-GS & P & N/A & P & P & P & P & N/A & P \\
49 & Local-GS & P & N/A & P & P & P & P & N/A & N/A \\
50 & Differentiable Hardware Rasterization & P & N/A & P & P & P & P & U & P \\
51 & Hardware-Rasterized Ray-Based Gaussian Splatting & P & N/A & U & P & P & P & U & N/A \\
52 & STREAMINGGS & P & P & P & P & P & P & P & N/A \\
53 & GCC & P & N/A & P & P & P & P & P & N/A \\
54 & LS-Gaussian & P & N/A & P & P & P & P & U & N/A \\
55 & 3DGauCIM & P & N/A & P & P & P & P & P & N/A \\
56 & PUP 3D-GS & P & P & U & U & P & U & N/A & U \\
57 & Speedy-Splat & P & P & P & U & P & R & N/A & N/A \\
58 & Splatonic & P & N/A & P & P & P & P & P & P \\
59 & Faster-GS & P & N/A & R & P & P & P & N/A & P \\
60 & Splaxel & P & N/A & P & P & P & P & N/A & P \\
61 & TideGS & P & N/A & P & U & P & P & N/A & P \\
62 & Nebula & P & N/A & P & P & P & P & P & N/A \\
\end{longtable}
\endgroup

\section{Reproducibility Protocol, GSPROF Toolkit, and Measurement Scope}
\label{app:reproducibility}
\label{app:arxiv-S2}

\subsection{Cross-Paper Evidence Admission}
\label{app:evidence-admission}

In the main article, Figs.~4 and~6(d) rely strictly on within-source comparisons without cross-paper normalization. A quantitative point is admitted only when its source specifies the affected workload or resource quantity, baseline and optimized conditions, input or aggregation scope, quality settings, and timed operations. Directly reported values, quantities derived arithmetically from reported table entries, and survey reproductions are tracked separately, while missing workload counters remain unavailable. Admission is independent of speedup direction, and the resulting points support neither cross-paper ranking nor regression.

In Fig.~4(a) and Fig.~4(b), RMF denotes Reducing the Memory Footprint, Light denotes LightGaussian, and Speedy denotes Speedy-Splat. C3DGS, C3DGS-B, and C3DGS-P label Compact 3DGS configurations, PUP and PUP+V denote PUP 3D-GS and PUP+Vectree, HAC-L and HAC-H identify HAC++ variants, Flux, Flux-D, and Flux-P correspond to Flux-GS, and Mobile-NQ and Mobile-tau.2 correspond to Mobile-GS. In Fig.~4(c), FLoD-543 and FLoD-432 denote FLoD detail-level sets $5/4/3$ and $4/3/2$, while Hier-tau2 and Hier-tau3 denote Hierarchical 3DGS settings. In Fig.~4(d), AdRGS denotes the AdRGS-AABB configuration of AdR-Gaussian~\cite{ref100}, using FPS and association counts from FlashGS~\cite{ref38} Tables~1 and~3, alongside gsplat and two FlashGS settings, FGS-Pre and FGS-Full. Every point preserves its provenance, reported or derived status, baseline, dataset or scene, aggregation scope, platform, backend, resolution, timing boundary, and quality context. Fig.~6(d) admits FlashGS, HiGS, Local-GS, gsplat, CLM, and Faster-GS under the same criteria, and aggregated points preserve matched operations and available quality differences.

\subsection{Controlled GPU Profiling}

All controlled GPU measurements run on identical NVIDIA RTX~3090 GPUs, with every paired baseline and variant executed on the same card. The tile sweep in Fig.~6(a)--(c) evaluates four scenes using 30K-iteration checkpoints. Bicycle and Bonsai from Mip-NeRF~360 contain 6,131,954 and 1,244,819 resident Gaussians at $800\times 532$ and $800\times 533$, respectively. Truck from Tanks \& Temples contains 2,541,226 resident Gaussians at $800\times 446$, and DrJohnson from Deep Blending contains 3,405,153 resident Gaussians at $800\times 526$. Each scene contributes 16 views across square tile widths of 8, 12, 16, 24, and 32 pixels, yielding 320 configuration--view measurements. Workload counts are extracted directly from the materialized Gaussian--tile and per-pixel candidate lists. Each scene--view pair is normalized to its 16~px tile baseline after verifying output parity and workload-count invariants.

Fig.~7 evaluates the same four scenes and 16 views per scene across five configurations, producing 320 measurements per panel. Relative to the width-800, 30K-checkpoint, 16~px-tile baseline, the other four configurations each vary a single factor. The 7K checkpoints contain 3,616,103, 1,157,141, 1,732,378, and 1,902,253 resident Gaussians for Bicycle, Bonsai, Truck, and DrJohnson, respectively. The higher-resolution setting uses $1600\times 1063$, $1600\times 1066$, $979\times 546$, and $1332\times 876$ in the same scene order. The remaining two configurations use tile widths of 8 and 32 pixels. The reported $R^{2}$ values come from ordinary least-squares fits in base-10 log--log coordinates, and $\rho$ denotes the Spearman rank correlation over the same 320 observations.

Fig.~8 and Table~\ref{tab:backend-supplement} evaluate Graphdeco, gsplat, FlashGS, StopThePop, Speedy-Splat, and Faster-GS on the 30K Truck checkpoint with 2,541,226 resident Gaussians at $800\times 446$, tile width 16, and four fixed views (IDs~0--3). Complete Forward latency is timed without profiler attachment, whereas kernel composition and DRAM traffic are gathered via separate Nsight Compute replays. Workload counts reflect backend-specific materialized work because culling and intersection semantics differ across renderers.

The extended suites use four RTX~3090 GPUs. GPU~0 times the five-anchor SH layouts and the two-width tile sweep. GPU~1 measures physical DRAM reads and writes for the 40 SH conditions via Nsight Compute application replay with warmed caches. GPU~2 evaluates the five rendering modes across the 127 test photographs and times the 12 camera routes at both widths. GPU~3 times the six backends across 127 views and measures physical DRAM traffic for 16 matched Graphdeco--gsplat view pairs via Nsight Compute kernel replay with cache flushing. The SH0-versus-SH3 controls run on GPU~0 for Bicycle and Truck and on GPU~2 for Bonsai and DrJohnson, while the sweeps in Figs.~6--8 each run on a single RTX~3090. Physical DRAM traffic sums all bytes read and written during one complete Forward pass. Forward latencies are always measured without profiler attachment, and diagnostic kernel intervals come from separate Nsight Systems replays. The attribute-layout suite uses PyTorch~2.6.0, CUDA~11.8, and gsplat~1.5.3. Peak memory allocations include resident tensors, and all timings cover monocular Forward rendering through image output while excluding compositor and display overhead.

\subsection{Reproducibility Boundary}

The accompanying artifact package provides figure inputs, provenance records, collection and plotting scripts, and measurement configurations, explicitly separating paper-reported, derived, and survey-measured values. Offline figure regeneration requires no GPU. Instrumentation defines $N_{\text{active}}$ as post-projection Gaussians that emit at least one materialized association and $N_{\text{pair}}$ as the materialized association-list length. Re-running GPU measurements requires the documented checkpoints, builds, and profiler settings and covers the measured rendering calls rather than end-to-end scene training.

\subsection{Open-Source Artifact}

The \href{https://github.com/HAPPYPMN/gsprof}{GSPROF repository} provides the code and measurements for main Table~5 and Secs.~S3--S4. From the root of the paper reproduction checkout, run \mbox{\texttt{python -m pip install -e .}} followed by \mbox{\texttt{gsprof paper}} to reproduce the tables on a CPU. The repository documents the measurement suites, GPU collection, and hardware-counter replay. GSPROF is MIT-licensed. Third-party components retain their respective licenses.

\section{Supplementary Backend Measurements for Fig.~8}
\label{app:backend-supplement}
\label{app:arxiv-S3}

Table~\ref{tab:backend-supplement} summarizes memory allocation and reconstruction fidelity for the matched-backend comparison in Fig.~8. Peak allocated memory covers the Gaussian representation, constructed association buffers, and temporary storage. PSNR, SSIM, and LPIPS are evaluated against the same ground-truth images, and each reported score difference is the maximum absolute per-view deviation from Graphdeco across the measured views.

\begin{table}[H]
\caption{Backend Memory and Reconstruction Quality on Fig.~8 Views}
\label{tab:backend-supplement}
\centering
\footnotesize
\setlength{\tabcolsep}{6pt}
\renewcommand{\arraystretch}{1.35}
\begin{tabular*}{\linewidth}{@{\extracolsep{\fill}}lcccc@{}}
\toprule
\multirow{2}{*}{Backend} & \multirow{2}{*}{\shortstack{Peak memory (GiB)\\median [min--max]}} & \multicolumn{3}{c}{Max. absolute score difference} \\
\cmidrule(lr){3-5}
& & PSNR & SSIM & LPIPS \\
\midrule
Graphdeco    & 1.92 [1.88, 1.94] & 0.000 & 0.0000 & 0.0000 \\
gsplat       & 0.97 [0.81, 0.99] & 0.036 & 0.0005 & 0.0003 \\
FlashGS      & 0.94 [0.94, 0.94] & 0.076 & 0.0028 & 0.0043 \\
StopThePop   & 1.96 [1.92, 1.98] & 0.001 & 0.0000 & 0.0000 \\
Speedy-Splat & 1.79 [1.75, 1.79] & 0.001 & 0.0000 & 0.0000 \\
Faster-GS    & 1.95 [1.79, 1.96] & 0.106 & 0.0045 & 0.0025 \\
\bottomrule
\end{tabular*}
\par\vspace{8pt}
\parbox{\linewidth}{\footnotesize Peak memory is summarized over measured views. Score differences use Graphdeco as the reference.}
\end{table}

At width 800, Table~\ref{tab:supp-six-extended} pairs unprofiled Forward latency with ground-truth image quality across 127 test photographs spanning all four scenes. On Bicycle, the median paired FlashGS PSNR difference relative to Graphdeco is $-0.126$~dB. Across four views per scene selected at the two endpoints and the two one-third positions of each test sequence, matched hardware-counter replays confirm that gsplat transfers more total DRAM bytes than Graphdeco on all four views of Bicycle, Bonsai, and Truck, and fewer on all four views of DrJohnson.

\begin{table}[H]
\caption{Forward Latency and Ground-Truth Image Quality on 127 Test Photographs}
\label{tab:supp-six-extended}
\centering\footnotesize\setlength{\tabcolsep}{3.5pt}\renewcommand{\arraystretch}{1.12}
\begin{tabular*}{\linewidth}{@{\extracolsep{\fill}}lrrrrrrrr@{}}
\toprule
Backend & \multicolumn{4}{c}{Bicycle (25 views)} & \multicolumn{4}{c}{Bonsai (37 views)} \\
\cmidrule(lr){2-5}\cmidrule(lr){6-9}
 & ms & PSNR & SSIM & LPIPS & ms & PSNR & SSIM & LPIPS \\
\midrule
Graphdeco & 7.00 & 24.28 & 0.797 & 0.167 & 2.35 & 29.75 & 0.941 & 0.076 \\
gsplat & 6.48 & 24.28 & 0.797 & 0.167 & 2.40 & 29.72 & 0.941 & 0.076 \\
FlashGS & 4.00 & 23.88 & 0.794 & 0.175 & 1.53 & 29.70 & 0.940 & 0.078 \\
StopThePop & 7.04 & 24.28 & 0.797 & 0.167 & 2.43 & 29.75 & 0.941 & 0.076 \\
Speedy-Splat & 4.79 & 24.28 & 0.797 & 0.167 & 1.86 & 29.75 & 0.941 & 0.076 \\
Faster-GS & 3.94 & 24.28 & 0.797 & 0.167 & 1.27 & 29.75 & 0.941 & 0.076 \\
\midrule
 & \multicolumn{4}{c}{Truck (32 views)} & \multicolumn{4}{c}{DrJohnson (33 views)} \\
\cmidrule(lr){2-5}\cmidrule(lr){6-9}
 & ms & PSNR & SSIM & LPIPS & ms & PSNR & SSIM & LPIPS \\
Graphdeco & 4.18 & 25.23 & 0.887 & 0.086 & 3.26 & 29.50 & 0.894 & 0.133 \\
gsplat & 3.79 & 25.20 & 0.886 & 0.085 & 2.20 & 29.50 & 0.894 & 0.133 \\
FlashGS & 2.39 & 25.19 & 0.886 & 0.089 & 1.06 & 29.37 & 0.895 & 0.134 \\
StopThePop & 4.18 & 25.23 & 0.887 & 0.086 & 3.33 & 29.50 & 0.894 & 0.133 \\
Speedy-Splat & 2.86 & 25.23 & 0.887 & 0.086 & 1.65 & 29.50 & 0.894 & 0.133 \\
Faster-GS & 2.18 & 25.22 & 0.887 & 0.086 & 1.26 & 29.50 & 0.894 & 0.133 \\
\bottomrule
\end{tabular*}
\par\smallskip\raggedright Latency is the median across photographs of their 30-call median Forward times (ms). PSNR (dB), SSIM, and LPIPS are scene medians against the same ground-truth photos. FlashGS reserves $32\times2^{20}$ association entries.
\end{table}

\section{Extended Controlled Profiling of Attribute Layouts, Tiling, Trajectories, and Defaults}
\label{app:extended-profiling}

\subsection{Full-SH3 Materialization Ablation}
\label{app:sh3-materialization}
\label{app:arxiv-S4.1}

All compared gsplat layouts hold scene geometry, third-order spherical harmonics, and classic alpha compositing fixed. Standard packed execution (P) copies visible SH coefficients into a compact temporary array. Indexed execution (I) reads resident coefficients directly by Gaussian identifier during shading. Native dense execution uses \texttt{packed=False} to operate directly on the full resident attribute arrays without constructing compact visible-subset buffers. An intermediate variant (M) materializes the packed SH array as in P while sharing the fused direction, mask, and SH shading kernel of I, isolating coefficient copying and addressing from kernel fusion. Active Gaussian and association counts agree identically across all 776 pose settings spanning four scenes and two widths, and the maximum absolute RGB channel difference between packed and indexed outputs is $4.77 \times 10^{-7}$.

On Bicycle, which contains 6.132 million Gaussians, native dense execution requires $5.01\times$ and $3.12\times$ the median temporary peak memory of indexed execution at widths 800 and 1920, respectively, while raising Forward P95 by roughly 0.5~ms at both widths. On Bonsai at width 1920 and on DrJohnson at both widths, native dense execution achieves a slightly lower P95 at a higher total peak allocation. Indexed execution attains the lowest total peak allocation across all eight configurations and improves both P95 and total peak relative to native dense execution in five configurations, while native dense execution can also reduce total peak allocation relative to packed execution.

For memory replays, PyTorch peak-allocation counters are reset immediately before each call and each rendered output is released before the next call. Temporary peak memory is reported as the median of the 97 per-pose increments, whereas total peak memory is the maximum allocation across those poses including resident tensors and camera state. Model loading, quality-metric networks, reference images, allocator reservations, and the CUDA context are excluded.

Relative to P, variant I eliminates three of four generic indexing kernel launches, although its SH kernel runs slightly longer in all eight configurations. Projection, Sorting, and Blending preserve identical kernel launch shapes and counts. For each configuration, photo anchor zero is replayed five times under Nsight Systems with CUDA and NVTX tracing, mapping each GPU kernel to its CPU launch and NVTX range via its CUDA correlation identifier. Here, Gather includes generic indexing inside view preparation, Sorting covers tile intersections, list ordering, and range encoding, and Blending covers pixel rasterization. Complete Forward latencies for causal attribution are collected in a separate process using 30 randomly interleaved calls per variant. Table~\ref{tab:supp-sh3-replays} reports stage call counts and summed kernel times in milliseconds, where SH copy is a subset of Gather and the final column sums all kernel intervals.

\begingroup\footnotesize\setlength{\tabcolsep}{3pt}
\begin{longtable}{lrcrrrrrrrrrr}
\caption{Complete Nsight Systems Replays at Photo Anchor Zero}\label{tab:supp-sh3-replays}\\
\toprule
Scene & Width & Mode & Replay & Gather & View & SH & Projection & Sorting & Blending & Other & SH copy & Sum \\
\midrule\endfirsthead
\toprule
Scene & Width & Mode & Replay & Gather & View & SH & Projection & Sorting & Blending & Other & SH copy & Sum \\
\midrule\endhead
\bottomrule\endfoot
Bicycle & 800 & P & 1 & 4, 1.272 & 14, 0.174 & 1, 0.458 & 4, 0.713 & 14, 1.088 & 1, 4.127 & 4, 0.191 & 1.023 & 8.022 \\
Bicycle & 800 & P & 2 & 4, 1.271 & 14, 0.174 & 1, 0.458 & 4, 0.711 & 14, 1.092 & 1, 4.095 & 4, 0.192 & 1.023 & 7.992 \\
Bicycle & 800 & P & 3 & 4, 1.271 & 14, 0.174 & 1, 0.459 & 4, 0.712 & 14, 1.085 & 1, 4.428 & 4, 0.191 & 1.023 & 8.320 \\
Bicycle & 800 & P & 4 & 4, 1.271 & 14, 0.174 & 1, 0.458 & 4, 0.711 & 14, 1.090 & 1, 4.173 & 4, 0.190 & 1.023 & 8.069 \\
Bicycle & 800 & P & 5 & 4, 1.272 & 14, 0.174 & 1, 0.459 & 4, 0.711 & 14, 1.092 & 1, 4.059 & 4, 0.191 & 1.023 & 7.958 \\
Bicycle & 800 & M & 1 & 2, 1.091 & 11, 0.023 & 1, 0.509 & 4, 0.711 & 14, 1.086 & 1, 4.328 & 4, 0.191 & 1.023 & 7.939 \\
Bicycle & 800 & M & 2 & 2, 1.091 & 11, 0.023 & 1, 0.510 & 4, 0.711 & 14, 1.089 & 1, 4.143 & 4, 0.191 & 1.023 & 7.758 \\
Bicycle & 800 & M & 3 & 2, 1.091 & 11, 0.023 & 1, 0.510 & 4, 0.712 & 14, 1.090 & 1, 4.164 & 4, 0.192 & 1.023 & 7.783 \\
Bicycle & 800 & M & 4 & 2, 1.091 & 11, 0.023 & 1, 0.510 & 4, 0.711 & 14, 1.090 & 1, 4.100 & 4, 0.191 & 1.023 & 7.717 \\
Bicycle & 800 & M & 5 & 2, 1.056 & 11, 0.021 & 1, 0.509 & 4, 0.708 & 14, 1.079 & 1, 3.724 & 4, 0.190 & 0.988 & 7.288 \\
Bicycle & 800 & I & 1 & 1, 0.068 & 11, 0.023 & 1, 0.511 & 4, 0.711 & 14, 1.086 & 1, 4.136 & 4, 0.191 & 0.000 & 6.727 \\
Bicycle & 800 & I & 2 & 1, 0.068 & 11, 0.023 & 1, 0.511 & 4, 0.710 & 14, 1.091 & 1, 4.339 & 4, 0.191 & 0.000 & 6.933 \\
Bicycle & 800 & I & 3 & 1, 0.069 & 11, 0.023 & 1, 0.511 & 4, 0.711 & 14, 1.091 & 1, 4.038 & 4, 0.192 & 0.000 & 6.634 \\
Bicycle & 800 & I & 4 & 1, 0.068 & 11, 0.023 & 1, 0.511 & 4, 0.711 & 14, 1.088 & 1, 4.201 & 4, 0.191 & 0.000 & 6.795 \\
Bicycle & 800 & I & 5 & 1, 0.068 & 11, 0.021 & 1, 0.510 & 4, 0.708 & 14, 1.077 & 1, 3.882 & 4, 0.190 & 0.000 & 6.457 \\
Bicycle & 1920 & P & 1 & 4, 1.219 & 14, 0.165 & 1, 0.458 & 4, 0.707 & 14, 1.774 & 1, 4.692 & 4, 0.190 & 0.976 & 9.206 \\
Bicycle & 1920 & P & 2 & 4, 1.220 & 14, 0.165 & 1, 0.457 & 4, 0.707 & 14, 1.781 & 1, 4.753 & 4, 0.190 & 0.976 & 9.274 \\
Bicycle & 1920 & P & 3 & 4, 1.241 & 14, 0.169 & 1, 0.458 & 4, 0.709 & 14, 1.796 & 1, 4.930 & 4, 0.190 & 0.996 & 9.493 \\
Bicycle & 1920 & P & 4 & 4, 1.238 & 14, 0.169 & 1, 0.458 & 4, 0.708 & 14, 1.794 & 1, 5.006 & 4, 0.190 & 0.992 & 9.562 \\
Bicycle & 1920 & P & 5 & 4, 1.237 & 14, 0.169 & 1, 0.458 & 4, 0.709 & 14, 1.794 & 1, 4.909 & 4, 0.190 & 0.993 & 9.466 \\
Bicycle & 1920 & M & 1 & 2, 1.044 & 11, 0.021 & 1, 0.508 & 4, 0.707 & 14, 1.774 & 1, 4.636 & 4, 0.190 & 0.976 & 8.880 \\
Bicycle & 1920 & M & 2 & 2, 1.064 & 11, 0.022 & 1, 0.509 & 4, 0.708 & 14, 1.799 & 1, 4.943 & 4, 0.191 & 0.996 & 9.236 \\
Bicycle & 1920 & M & 3 & 2, 1.062 & 11, 0.022 & 1, 0.509 & 4, 0.708 & 14, 1.793 & 1, 4.986 & 4, 0.190 & 0.993 & 9.269 \\
Bicycle & 1920 & M & 4 & 2, 1.062 & 11, 0.022 & 1, 0.508 & 4, 0.707 & 14, 1.788 & 1, 4.924 & 4, 0.190 & 0.993 & 9.201 \\
Bicycle & 1920 & M & 5 & 2, 1.061 & 11, 0.022 & 1, 0.509 & 4, 0.707 & 14, 1.793 & 1, 4.936 & 4, 0.190 & 0.993 & 9.216 \\
Bicycle & 1920 & I & 1 & 1, 0.068 & 11, 0.020 & 1, 0.509 & 4, 0.709 & 14, 1.778 & 1, 4.648 & 4, 0.190 & 0.000 & 7.922 \\
Bicycle & 1920 & I & 2 & 1, 0.068 & 11, 0.021 & 1, 0.510 & 4, 0.708 & 14, 1.772 & 1, 4.995 & 4, 0.188 & 0.000 & 8.262 \\
Bicycle & 1920 & I & 3 & 1, 0.069 & 11, 0.022 & 1, 0.510 & 4, 0.708 & 14, 1.784 & 1, 4.897 & 4, 0.189 & 0.000 & 8.179 \\
Bicycle & 1920 & I & 4 & 1, 0.068 & 11, 0.022 & 1, 0.510 & 4, 0.707 & 14, 1.786 & 1, 4.942 & 4, 0.190 & 0.000 & 8.225 \\
Bicycle & 1920 & I & 5 & 1, 0.068 & 11, 0.022 & 1, 0.509 & 4, 0.710 & 14, 1.784 & 1, 4.931 & 4, 0.190 & 0.000 & 8.214 \\
Bonsai & 800 & P & 1 & 4, 0.303 & 14, 0.061 & 1, 0.110 & 4, 0.158 & 14, 0.390 & 1, 1.501 & 4, 0.037 & 0.238 & 2.560 \\
Bonsai & 800 & P & 2 & 4, 0.305 & 14, 0.061 & 1, 0.110 & 4, 0.158 & 14, 0.396 & 1, 1.562 & 4, 0.037 & 0.238 & 2.628 \\
Bonsai & 800 & P & 3 & 4, 0.305 & 14, 0.061 & 1, 0.109 & 4, 0.159 & 14, 0.397 & 1, 1.614 & 4, 0.036 & 0.238 & 2.680 \\
Bonsai & 800 & P & 4 & 4, 0.305 & 14, 0.061 & 1, 0.109 & 4, 0.158 & 14, 0.395 & 1, 1.607 & 4, 0.037 & 0.238 & 2.672 \\
Bonsai & 800 & P & 5 & 4, 0.305 & 14, 0.060 & 1, 0.109 & 4, 0.158 & 14, 0.393 & 1, 1.556 & 4, 0.037 & 0.238 & 2.618 \\
Bonsai & 800 & M & 1 & 2, 0.258 & 11, 0.023 & 1, 0.121 & 4, 0.158 & 14, 0.399 & 1, 1.522 & 4, 0.036 & 0.239 & 2.517 \\
Bonsai & 800 & M & 2 & 2, 0.257 & 11, 0.023 & 1, 0.121 & 4, 0.159 & 14, 0.389 & 1, 1.641 & 4, 0.037 & 0.239 & 2.627 \\
Bonsai & 800 & M & 3 & 2, 0.257 & 11, 0.023 & 1, 0.121 & 4, 0.160 & 14, 0.397 & 1, 1.512 & 4, 0.036 & 0.239 & 2.506 \\
Bonsai & 800 & M & 4 & 2, 0.258 & 11, 0.023 & 1, 0.121 & 4, 0.158 & 14, 0.394 & 1, 1.566 & 4, 0.037 & 0.239 & 2.556 \\
Bonsai & 800 & M & 5 & 2, 0.257 & 11, 0.023 & 1, 0.121 & 4, 0.158 & 14, 0.392 & 1, 1.530 & 4, 0.037 & 0.239 & 2.518 \\
Bonsai & 800 & I & 1 & 1, 0.019 & 11, 0.023 & 1, 0.121 & 4, 0.158 & 14, 0.398 & 1, 1.508 & 4, 0.036 & 0.000 & 2.264 \\
Bonsai & 800 & I & 2 & 1, 0.019 & 11, 0.023 & 1, 0.121 & 4, 0.158 & 14, 0.390 & 1, 1.564 & 4, 0.037 & 0.000 & 2.312 \\
Bonsai & 800 & I & 3 & 1, 0.019 & 11, 0.023 & 1, 0.121 & 4, 0.158 & 14, 0.399 & 1, 1.557 & 4, 0.036 & 0.000 & 2.313 \\
Bonsai & 800 & I & 4 & 1, 0.019 & 11, 0.023 & 1, 0.121 & 4, 0.159 & 14, 0.399 & 1, 1.506 & 4, 0.036 & 0.000 & 2.262 \\
Bonsai & 800 & I & 5 & 1, 0.019 & 11, 0.023 & 1, 0.121 & 4, 0.158 & 14, 0.394 & 1, 1.532 & 4, 0.036 & 0.000 & 2.283 \\
Bonsai & 1920 & P & 1 & 4, 0.304 & 14, 0.061 & 1, 0.109 & 4, 0.160 & 14, 0.950 & 1, 2.110 & 4, 0.036 & 0.238 & 3.729 \\
Bonsai & 1920 & P & 2 & 4, 0.304 & 14, 0.061 & 1, 0.109 & 4, 0.161 & 14, 0.965 & 1, 2.115 & 4, 0.037 & 0.238 & 3.751 \\
Bonsai & 1920 & P & 3 & 4, 0.304 & 14, 0.061 & 1, 0.109 & 4, 0.161 & 14, 0.920 & 1, 2.113 & 4, 0.036 & 0.238 & 3.705 \\
Bonsai & 1920 & P & 4 & 4, 0.304 & 14, 0.061 & 1, 0.109 & 4, 0.161 & 14, 0.946 & 1, 2.121 & 4, 0.037 & 0.238 & 3.739 \\
Bonsai & 1920 & P & 5 & 4, 0.304 & 14, 0.061 & 1, 0.109 & 4, 0.161 & 14, 0.930 & 1, 2.108 & 4, 0.036 & 0.238 & 3.709 \\
Bonsai & 1920 & M & 1 & 2, 0.257 & 11, 0.023 & 1, 0.121 & 4, 0.161 & 14, 0.946 & 1, 2.106 & 4, 0.036 & 0.238 & 3.651 \\
Bonsai & 1920 & M & 2 & 2, 0.257 & 11, 0.023 & 1, 0.121 & 4, 0.161 & 14, 0.944 & 1, 2.118 & 4, 0.036 & 0.238 & 3.661 \\
Bonsai & 1920 & M & 3 & 2, 0.257 & 11, 0.023 & 1, 0.121 & 4, 0.161 & 14, 0.934 & 1, 2.116 & 4, 0.037 & 0.238 & 3.650 \\
Bonsai & 1920 & M & 4 & 2, 0.257 & 11, 0.023 & 1, 0.121 & 4, 0.161 & 14, 0.953 & 1, 2.122 & 4, 0.037 & 0.239 & 3.674 \\
Bonsai & 1920 & M & 5 & 2, 0.257 & 11, 0.023 & 1, 0.121 & 4, 0.161 & 14, 0.928 & 1, 2.105 & 4, 0.037 & 0.238 & 3.632 \\
Bonsai & 1920 & I & 1 & 1, 0.018 & 11, 0.023 & 1, 0.121 & 4, 0.161 & 14, 0.952 & 1, 2.117 & 4, 0.037 & 0.000 & 3.430 \\
Bonsai & 1920 & I & 2 & 1, 0.019 & 11, 0.023 & 1, 0.121 & 4, 0.161 & 14, 0.933 & 1, 2.128 & 4, 0.037 & 0.000 & 3.422 \\
Bonsai & 1920 & I & 3 & 1, 0.019 & 11, 0.023 & 1, 0.121 & 4, 0.162 & 14, 0.957 & 1, 2.116 & 4, 0.036 & 0.000 & 3.435 \\
Bonsai & 1920 & I & 4 & 1, 0.019 & 11, 0.023 & 1, 0.121 & 4, 0.161 & 14, 0.937 & 1, 2.127 & 4, 0.037 & 0.000 & 3.425 \\
Bonsai & 1920 & I & 5 & 1, 0.019 & 11, 0.023 & 1, 0.121 & 4, 0.160 & 14, 0.956 & 1, 2.122 & 4, 0.036 & 0.000 & 3.437 \\
Truck & 800 & P & 1 & 4, 0.707 & 14, 0.108 & 1, 0.255 & 4, 0.322 & 14, 0.668 & 1, 2.043 & 4, 0.107 & 0.564 & 4.212 \\
Truck & 800 & P & 2 & 4, 0.707 & 14, 0.108 & 1, 0.256 & 4, 0.323 & 14, 0.666 & 1, 1.981 & 4, 0.107 & 0.564 & 4.148 \\
Truck & 800 & P & 3 & 4, 0.707 & 14, 0.109 & 1, 0.255 & 4, 0.324 & 14, 0.665 & 1, 2.018 & 4, 0.107 & 0.564 & 4.186 \\
Truck & 800 & P & 4 & 4, 0.707 & 14, 0.109 & 1, 0.255 & 4, 0.324 & 14, 0.667 & 1, 2.122 & 4, 0.107 & 0.564 & 4.291 \\
Truck & 800 & P & 5 & 4, 0.706 & 14, 0.108 & 1, 0.255 & 4, 0.324 & 14, 0.668 & 1, 2.019 & 4, 0.106 & 0.564 & 4.187 \\
Truck & 800 & M & 1 & 2, 0.605 & 11, 0.023 & 1, 0.283 & 4, 0.323 & 14, 0.668 & 1, 2.037 & 4, 0.107 & 0.565 & 4.048 \\
Truck & 800 & M & 2 & 2, 0.605 & 11, 0.023 & 1, 0.283 & 4, 0.323 & 14, 0.664 & 1, 2.002 & 4, 0.107 & 0.565 & 4.007 \\
Truck & 800 & M & 3 & 2, 0.605 & 11, 0.023 & 1, 0.283 & 4, 0.324 & 14, 0.673 & 1, 2.034 & 4, 0.106 & 0.565 & 4.049 \\
Truck & 800 & M & 4 & 2, 0.605 & 11, 0.023 & 1, 0.283 & 4, 0.323 & 14, 0.670 & 1, 2.010 & 4, 0.107 & 0.565 & 4.021 \\
Truck & 800 & M & 5 & 2, 0.605 & 11, 0.023 & 1, 0.283 & 4, 0.323 & 14, 0.667 & 1, 2.016 & 4, 0.107 & 0.565 & 4.024 \\
Truck & 800 & I & 1 & 1, 0.040 & 11, 0.023 & 1, 0.284 & 4, 0.322 & 14, 0.664 & 1, 2.027 & 4, 0.107 & 0.000 & 3.468 \\
Truck & 800 & I & 2 & 1, 0.040 & 11, 0.023 & 1, 0.284 & 4, 0.323 & 14, 0.666 & 1, 2.039 & 4, 0.107 & 0.000 & 3.482 \\
Truck & 800 & I & 3 & 1, 0.040 & 11, 0.023 & 1, 0.284 & 4, 0.322 & 14, 0.666 & 1, 2.030 & 4, 0.107 & 0.000 & 3.473 \\
Truck & 800 & I & 4 & 1, 0.040 & 11, 0.023 & 1, 0.284 & 4, 0.323 & 14, 0.674 & 1, 2.048 & 4, 0.106 & 0.000 & 3.498 \\
Truck & 800 & I & 5 & 1, 0.040 & 11, 0.023 & 1, 0.284 & 4, 0.323 & 14, 0.669 & 1, 1.988 & 4, 0.106 & 0.000 & 3.433 \\
Truck & 1920 & P & 1 & 4, 0.705 & 14, 0.108 & 1, 0.255 & 4, 0.325 & 14, 1.261 & 1, 2.931 & 4, 0.107 & 0.563 & 5.692 \\
Truck & 1920 & P & 2 & 4, 0.705 & 14, 0.108 & 1, 0.255 & 4, 0.326 & 14, 1.271 & 1, 2.909 & 4, 0.107 & 0.563 & 5.681 \\
Truck & 1920 & P & 3 & 4, 0.705 & 14, 0.108 & 1, 0.255 & 4, 0.324 & 14, 1.262 & 1, 2.923 & 4, 0.107 & 0.563 & 5.683 \\
Truck & 1920 & P & 4 & 4, 0.704 & 14, 0.108 & 1, 0.255 & 4, 0.325 & 14, 1.259 & 1, 2.916 & 4, 0.107 & 0.563 & 5.674 \\
Truck & 1920 & P & 5 & 4, 0.705 & 14, 0.108 & 1, 0.255 & 4, 0.324 & 14, 1.267 & 1, 2.942 & 4, 0.107 & 0.563 & 5.707 \\
Truck & 1920 & M & 1 & 2, 0.604 & 11, 0.023 & 1, 0.283 & 4, 0.325 & 14, 1.273 & 1, 2.906 & 4, 0.106 & 0.564 & 5.520 \\
Truck & 1920 & M & 2 & 2, 0.603 & 11, 0.023 & 1, 0.283 & 4, 0.325 & 14, 1.263 & 1, 2.933 & 4, 0.107 & 0.563 & 5.538 \\
Truck & 1920 & M & 3 & 2, 0.603 & 11, 0.023 & 1, 0.283 & 4, 0.325 & 14, 1.265 & 1, 2.920 & 4, 0.107 & 0.563 & 5.526 \\
Truck & 1920 & M & 4 & 2, 0.603 & 11, 0.023 & 1, 0.283 & 4, 0.325 & 14, 1.273 & 1, 2.912 & 4, 0.107 & 0.563 & 5.526 \\
Truck & 1920 & M & 5 & 2, 0.603 & 11, 0.023 & 1, 0.283 & 4, 0.324 & 14, 1.276 & 1, 2.907 & 4, 0.106 & 0.563 & 5.523 \\
Truck & 1920 & I & 1 & 1, 0.040 & 11, 0.023 & 1, 0.283 & 4, 0.326 & 14, 1.269 & 1, 2.909 & 4, 0.106 & 0.000 & 4.956 \\
Truck & 1920 & I & 2 & 1, 0.040 & 11, 0.023 & 1, 0.283 & 4, 0.324 & 14, 1.275 & 1, 2.921 & 4, 0.106 & 0.000 & 4.972 \\
Truck & 1920 & I & 3 & 1, 0.040 & 11, 0.023 & 1, 0.283 & 4, 0.324 & 14, 1.263 & 1, 2.898 & 4, 0.107 & 0.000 & 4.938 \\
Truck & 1920 & I & 4 & 1, 0.040 & 11, 0.023 & 1, 0.283 & 4, 0.326 & 14, 1.259 & 1, 2.910 & 4, 0.106 & 0.000 & 4.947 \\
Truck & 1920 & I & 5 & 1, 0.040 & 11, 0.023 & 1, 0.283 & 4, 0.325 & 14, 1.264 & 1, 2.900 & 4, 0.107 & 0.000 & 4.944 \\
DrJohnson & 800 & P & 1 & 4, 0.187 & 14, 0.043 & 1, 0.062 & 4, 0.364 & 14, 0.322 & 1, 0.976 & 4, 0.018 & 0.134 & 1.973 \\
DrJohnson & 800 & P & 2 & 4, 0.187 & 14, 0.043 & 1, 0.062 & 4, 0.365 & 14, 0.323 & 1, 1.021 & 4, 0.018 & 0.134 & 2.018 \\
DrJohnson & 800 & P & 3 & 4, 0.187 & 14, 0.044 & 1, 0.062 & 4, 0.365 & 14, 0.329 & 1, 1.000 & 4, 0.018 & 0.134 & 2.003 \\
DrJohnson & 800 & P & 4 & 4, 0.188 & 14, 0.043 & 1, 0.062 & 4, 0.364 & 14, 0.329 & 1, 0.988 & 4, 0.018 & 0.134 & 1.992 \\
DrJohnson & 800 & P & 5 & 4, 0.187 & 14, 0.043 & 1, 0.062 & 4, 0.364 & 14, 0.339 & 1, 1.035 & 4, 0.018 & 0.134 & 2.047 \\
DrJohnson & 800 & M & 1 & 2, 0.150 & 11, 0.023 & 1, 0.072 & 4, 0.364 & 14, 0.321 & 1, 1.032 & 4, 0.018 & 0.135 & 1.981 \\
DrJohnson & 800 & M & 2 & 2, 0.151 & 11, 0.023 & 1, 0.072 & 4, 0.363 & 14, 0.323 & 1, 0.990 & 4, 0.018 & 0.135 & 1.941 \\
DrJohnson & 800 & M & 3 & 2, 0.150 & 11, 0.023 & 1, 0.072 & 4, 0.365 & 14, 0.322 & 1, 1.015 & 4, 0.018 & 0.135 & 1.964 \\
DrJohnson & 800 & M & 4 & 2, 0.151 & 11, 0.023 & 1, 0.072 & 4, 0.364 & 14, 0.328 & 1, 0.974 & 4, 0.018 & 0.135 & 1.929 \\
DrJohnson & 800 & M & 5 & 2, 0.151 & 11, 0.023 & 1, 0.072 & 4, 0.364 & 14, 0.322 & 1, 0.991 & 4, 0.018 & 0.135 & 1.940 \\
DrJohnson & 800 & I & 1 & 1, 0.016 & 11, 0.023 & 1, 0.072 & 4, 0.364 & 14, 0.325 & 1, 0.990 & 4, 0.018 & 0.000 & 1.808 \\
DrJohnson & 800 & I & 2 & 1, 0.016 & 11, 0.023 & 1, 0.072 & 4, 0.364 & 14, 0.322 & 1, 1.026 & 4, 0.019 & 0.000 & 1.841 \\
DrJohnson & 800 & I & 3 & 1, 0.016 & 11, 0.022 & 1, 0.072 & 4, 0.364 & 14, 0.324 & 1, 1.008 & 4, 0.019 & 0.000 & 1.825 \\
DrJohnson & 800 & I & 4 & 1, 0.016 & 11, 0.023 & 1, 0.072 & 4, 0.364 & 14, 0.322 & 1, 1.008 & 4, 0.018 & 0.000 & 1.822 \\
DrJohnson & 800 & I & 5 & 1, 0.016 & 11, 0.023 & 1, 0.072 & 4, 0.363 & 14, 0.324 & 1, 1.031 & 4, 0.018 & 0.000 & 1.848 \\
DrJohnson & 1920 & P & 1 & 4, 0.186 & 14, 0.043 & 1, 0.062 & 4, 0.366 & 14, 0.861 & 1, 2.371 & 4, 0.018 & 0.133 & 3.907 \\
DrJohnson & 1920 & P & 2 & 4, 0.187 & 14, 0.043 & 1, 0.062 & 4, 0.366 & 14, 0.877 & 1, 2.383 & 4, 0.018 & 0.134 & 3.936 \\
DrJohnson & 1920 & P & 3 & 4, 0.187 & 14, 0.043 & 1, 0.062 & 4, 0.366 & 14, 0.869 & 1, 2.385 & 4, 0.018 & 0.133 & 3.929 \\
DrJohnson & 1920 & P & 4 & 4, 0.187 & 14, 0.043 & 1, 0.062 & 4, 0.365 & 14, 0.872 & 1, 2.396 & 4, 0.018 & 0.134 & 3.944 \\
DrJohnson & 1920 & P & 5 & 4, 0.186 & 14, 0.043 & 1, 0.061 & 4, 0.366 & 14, 0.868 & 1, 2.383 & 4, 0.018 & 0.134 & 3.926 \\
DrJohnson & 1920 & M & 1 & 2, 0.150 & 11, 0.023 & 1, 0.071 & 4, 0.366 & 14, 0.862 & 1, 2.385 & 4, 0.018 & 0.134 & 3.875 \\
DrJohnson & 1920 & M & 2 & 2, 0.150 & 11, 0.023 & 1, 0.071 & 4, 0.365 & 14, 0.864 & 1, 2.378 & 4, 0.018 & 0.134 & 3.870 \\
DrJohnson & 1920 & M & 3 & 2, 0.150 & 11, 0.023 & 1, 0.071 & 4, 0.366 & 14, 0.888 & 1, 2.386 & 4, 0.018 & 0.134 & 3.902 \\
DrJohnson & 1920 & M & 4 & 2, 0.150 & 11, 0.023 & 1, 0.071 & 4, 0.365 & 14, 0.862 & 1, 2.387 & 4, 0.018 & 0.134 & 3.876 \\
DrJohnson & 1920 & M & 5 & 2, 0.150 & 11, 0.023 & 1, 0.072 & 4, 0.365 & 14, 0.876 & 1, 2.372 & 4, 0.018 & 0.134 & 3.875 \\
DrJohnson & 1920 & I & 1 & 1, 0.016 & 11, 0.023 & 1, 0.072 & 4, 0.366 & 14, 0.868 & 1, 2.361 & 4, 0.018 & 0.000 & 3.724 \\
DrJohnson & 1920 & I & 2 & 1, 0.016 & 11, 0.023 & 1, 0.072 & 4, 0.365 & 14, 0.859 & 1, 2.393 & 4, 0.018 & 0.000 & 3.746 \\
DrJohnson & 1920 & I & 3 & 1, 0.016 & 11, 0.023 & 1, 0.072 & 4, 0.365 & 14, 0.863 & 1, 2.394 & 4, 0.018 & 0.000 & 3.751 \\
DrJohnson & 1920 & I & 4 & 1, 0.016 & 11, 0.023 & 1, 0.072 & 4, 0.366 & 14, 0.863 & 1, 2.386 & 4, 0.018 & 0.000 & 3.743 \\
DrJohnson & 1920 & I & 5 & 1, 0.016 & 11, 0.023 & 1, 0.072 & 4, 0.366 & 14, 0.891 & 1, 2.384 & 4, 0.018 & 0.000 & 3.769 \\
\end{longtable}
\endgroup

Expanding the matched ablation to five photo anchors per scene at both widths shows that indexed execution saves 0.085--1.204~ms over packed execution across all 40 conditions. Computing $(t_{M}-t_{I})/(t_{P}-t_{I})$ per anchor yields a materialization share of 24.5--83.1\%, with a median of 74.8\%. DrJohnson spans the full percentage range at width 800, where its packed-to-indexed savings range from 0.104 to 1.080~ms, as summarized in Table~\ref{tab:supp-sh-traffic}. Independent Nsight Compute replays confirm that variant I reduces total physical DRAM reads and writes by 6.7--23.7\% relative to M at a median of 17.9\%, and by 8.0--28.0\% relative to P.

\begin{table}[H]
\caption{Five-Anchor SH Timing and Physical DRAM Traffic}
\label{tab:supp-sh-traffic}
\centering\footnotesize\setlength{\tabcolsep}{3.5pt}\renewcommand{\arraystretch}{1.12}
\begin{tabular*}{\linewidth}{@{\extracolsep{\fill}}lrrrrrr@{}}
\toprule
& & & & & \multicolumn{2}{c}{DRAM reduction (\%)} \\
\cmidrule(lr){6-7}
Scene & Width & $t_P-t_I$ (ms) & $t_M-t_I$ (ms) & Share (\%) & vs.\ $M$ & vs.\ $P$ \\
\midrule
Bicycle & 800 & 0.822--1.204 & 0.635--0.939 & 73.1--78.0 & 20.1--22.2 & 23.9--26.3 \\
Bicycle & 1920 & 0.822--1.194 & 0.634--0.915 & 75.9--78.6 & 15.9--18.2 & 19.1--21.8 \\
Bonsai & 800 & 0.238--0.287 & 0.166--0.224 & 67.4--81.9 & 21.8--22.7 & 25.8--26.8 \\
Bonsai & 1920 & 0.178--0.280 & 0.090--0.207 & 50.6--75.0 & 14.7--16.1 & 17.7--19.3 \\
Truck & 800 & 0.366--0.647 & 0.243--0.498 & 66.1--78.1 & 19.9--23.1 & 23.6--27.3 \\
Truck & 1920 & 0.321--0.650 & 0.235--0.483 & 73.2--75.5 & 14.0--18.3 & 16.9--21.9 \\
DrJohnson & 800 & 0.104--1.080 & 0.026--0.839 & 24.5--83.1 & 11.0--23.7 & 13.0--28.0 \\
DrJohnson & 1920 & 0.085--1.064 & 0.025--0.808 & 28.9--75.9 & 6.7--18.2 & 8.0--21.9 \\
\bottomrule
\end{tabular*}
\par\smallskip\raggedright Each range spans the five matched photo anchors in that row. Times are medians of 30 unprofiled Forward calls per variant. The last two columns give per-anchor percentage reductions in physical DRAM reads plus writes for $I$.
\end{table}

Across the same 40 scene, width, and anchor conditions, comparing SH0, which stores one RGB triplet or 12~B per Gaussian, against SH3, which stores 16 triplets or 192~B per Gaussian, confirms that the packed-to-indexed time saving is larger for SH3 in every condition by 0.029--0.881~ms. Because this endpoint control alters both attribute width and shading arithmetic, it serves as diagnostic evidence that materialization cost scales with per-Gaussian payload size.

\subsection{CompGS Representation Lifecycle}
\label{app:compgs-lifecycle}
\label{app:arxiv-S4.2}

We evaluate the two released CompGS Truck codebooks~\cite{compgsVQ} with 4k and 32k entries, which occupy 70.12 and 71.61~MB on disk, 83.06 and 84.55~MB as indexed GPU tensors, and 611.00~MB each once expanded. Rendering all 251 test views at widths 800 and 1920, we compute for each camera the median GPU time difference between per-frame expansion and load-time decoding. The median paired overhead is 3.416~ms at width 800 and 3.406~ms at width 1920 for the 4k codebook, and 3.456 and 3.446~ms for the 32k codebook, with both decoding schedules producing identical output pixels on every view. At width 800, per-frame expansion increases the scene-median frame time by $1.90\times$ for both codebooks.

\subsection{Tile Granularity, Image Quality, and Trajectory Budgets}
\label{app:trajectory-budgets}
\label{app:arxiv-S4.3}

Across the four 30K checkpoints, we sweep five tile widths at image widths of 800 and 1600 pixels over 16 views per scene, normalizing each view to its own 16~px tile latency before taking the scene median in Table~\ref{tab:supp-tile}. An intermediate tile outperforms both 8 and 32~px endpoints on 127 of 128 paired views and widths. In the sole exception, DrJohnson view \texttt{IMG\_6349} at width 800, the 8~px tile takes 10.17~ms compared with 10.27~ms for the fastest intermediate width of 12~px.

\begin{table}[H]
\caption{Median Normalized Forward Time by Tile Width}
\label{tab:supp-tile}
\centering\footnotesize\setlength{\tabcolsep}{3.5pt}\renewcommand{\arraystretch}{1.12}
\begin{tabular*}{\linewidth}{@{\extracolsep{\fill}}lrrrrrr@{}}
\toprule
Scene & Image width & 8 px & 12 px & 16 px & 24 px & 32 px \\
\midrule
Bicycle & 800 & 1.136 & \textbf{0.975} & 1.000 & 1.222 & 1.525 \\
Bicycle & 1600 & 1.470 & 1.135 & \textbf{1.000} & 1.104 & 1.315 \\
Bonsai & 800 & 1.121 & \textbf{0.982} & 1.000 & 1.237 & 1.546 \\
Bonsai & 1600 & 1.574 & 1.167 & \textbf{1.000} & 1.047 & 1.208 \\
Truck & 800 & 1.166 & 1.002 & \textbf{1.000} & 1.199 & 1.537 \\
Truck & 1600 & 1.588 & 1.168 & \textbf{1.000} & 1.039 & 1.203 \\
DrJohnson & 800 & 1.299 & 1.067 & \textbf{1.000} & 1.134 & 1.293 \\
DrJohnson & 1600 & 1.635 & 1.209 & \textbf{1.000} & 1.005 & 1.098 \\
\bottomrule
\end{tabular*}
\par\smallskip\raggedright Each row summarizes 16 views. Bold indicates the minimum of the unrounded scene medians.
\end{table}

For visual-quality gating, all five modes render the same 127 test photographs at widths 800 and 1920 using classic compositing, third-order SH, a black background, and 16~px tiles. PSNR, SSIM, and AlexNet LPIPS are evaluated against each photograph. A mode passes when its PSNR and SSIM drops relative to packed execution do not exceed 0.2~dB and 0.005 and its LPIPS increase does not exceed 0.01. Radius clipping uses a 2~px threshold, and Table~\ref{tab:supp-clean-quality} reports the number of passing photographs per scene and width.

\begingroup
\captionsetup[table]{skip=4pt}
\begin{table}[H]
\caption{Test Photographs Meeting All Three Quality Bounds}
\label{tab:supp-clean-quality}
\centering\footnotesize\setlength{\tabcolsep}{3.5pt}\renewcommand{\arraystretch}{1.04}
\begin{tabular*}{\linewidth}{@{\extracolsep{\fill}}lrrrrrrr@{}}
\toprule
Scene & Width & Photos & Packed & Dense & Indexed & Clip & Clip + Indexed \\
\midrule
Bicycle & 800 & 25 & 25 & 25 & 25 & 4 & 4 \\
Bicycle & 1920 & 25 & 25 & 25 & 25 & 11 & 11 \\
Bonsai & 800 & 37 & 37 & 37 & 37 & 37 & 37 \\
Bonsai & 1920 & 37 & 37 & 37 & 37 & 37 & 37 \\
Truck & 800 & 32 & 32 & 32 & 32 & 19 & 19 \\
Truck & 1920 & 32 & 32 & 32 & 32 & 26 & 26 \\
DrJohnson & 800 & 33 & 33 & 33 & 33 & 33 & 33 \\
DrJohnson & 1920 & 33 & 33 & 33 & 33 & 33 & 33 \\
\bottomrule
\end{tabular*}
\par\smallskip\raggedright The same 127 distinct photographs appear at both widths. The totals passing with clipping are 93 at width 800 and 107 at width 1920, with or without indexed shading.
\end{table}

For trajectory timing, each scene provides three routes built from disjoint sets of five photo anchors: Route~0 uses test-camera indices 0, 3, 6, 9, and 12, while Routes~1 and~2 sample five evenly spaced indices from the two halves of the remaining test sequence starting at index 13. Linear center interpolation and spherical rotation interpolation generate 97 poses per route. After 12 warm-up calls per mode, the five modes are randomly shuffled within each pose and timed for 30 calls per mode via CUDA events. P50, P95, and P99 summarize the 97 per-pose medians, and any pose median exceeding $1000/120$~ms counts as a 120~Hz deadline miss in Table~\ref{tab:supp-clean-routes}. In addition, a diagnostic stage replay of Truck frame~60 at width 1920 measures 3.50~ms for Graphdeco radix Sorting versus 2.94~ms for Blending rasterization. All trajectory budgets use unprofiled Forward calls.

\begin{table}[H]
\caption{Five-Mode Trajectory Timing and Missed Frames at 120 Hz}
\label{tab:supp-clean-routes}
\centering\scriptsize\setlength{\tabcolsep}{2pt}\renewcommand{\arraystretch}{1.02}
\begin{tabular*}{\linewidth}{@{\extracolsep{\fill}}lrrccccc@{}}
\toprule
Scene & Route & Width & Packed & Dense & Indexed & Clip & Clip + Indexed \\
\midrule
Bicycle & 0 & 800 & 3.86/6.75/7.78/1 & 4.41/6.40/7.13/0 & 3.50/5.74/6.57/0 & 3.44/5.34/5.77/0 & 3.19/4.75/5.08/0 \\
 & 0 & 1920 & 6.63/9.42/10.24/15 & 7.02/8.98/9.51/12 & 6.18/8.43/9.04/6 & 6.36/8.81/9.33/7 & 6.07/7.96/8.36/2 \\
 & 1 & 800 & 6.35/7.30/7.44/0 & 6.01/7.15/7.26/0 & 5.39/6.47/6.65/0 & 4.99/5.65/5.93/0 & 4.38/5.09/5.37/0 \\
 & 1 & 1920 & 8.32/9.10/9.12/47 & 8.20/8.62/8.63/24 & 7.65/8.10/8.11/0 & 7.99/8.38/8.40/10 & 7.23/7.61/7.78/0 \\
 & 2 & 800 & 5.36/6.48/7.02/0 & 5.46/6.37/6.82/0 & 4.70/5.70/6.26/0 & 4.50/5.02/5.24/0 & 4.06/4.52/4.70/0 \\
 & 2 & 1920 & 7.64/8.50/8.65/10 & 7.51/8.38/8.60/7 & 6.88/7.74/7.95/0 & 7.21/8.11/8.36/3 & 6.58/7.50/7.77/0 \\
\midrule
Bonsai & 0 & 800 & 2.74/3.06/3.21/0 & 2.48/2.81/2.94/0 & 2.46/2.78/2.90/0 & 2.29/2.50/2.52/0 & 2.07/2.25/2.29/0 \\
 & 0 & 1920 & 4.00/4.16/4.18/0 & 3.68/3.80/3.82/0 & 3.71/3.85/3.86/0 & 3.94/4.08/4.10/0 & 3.67/3.79/3.82/0 \\
 & 1 & 800 & 2.54/2.71/2.76/0 & 2.29/2.47/2.51/0 & 2.28/2.42/2.48/0 & 2.11/2.27/2.28/0 & 1.90/2.04/2.06/0 \\
 & 1 & 1920 & 3.81/4.05/4.08/0 & 3.52/3.74/3.76/0 & 3.54/3.77/3.80/0 & 3.77/4.00/4.02/0 & 3.51/3.73/3.74/0 \\
 & 2 & 800 & 2.30/2.48/2.49/0 & 2.15/2.30/2.32/0 & 2.12/2.27/2.29/0 & 2.05/2.22/2.25/0 & 1.89/2.04/2.06/0 \\
 & 2 & 1920 & 3.72/3.98/3.99/0 & 3.52/3.71/3.72/0 & 3.53/3.75/3.76/0 & 3.69/3.93/3.94/0 & 3.50/3.71/3.72/0 \\
\midrule
Truck & 0 & 800 & 3.21/3.79/4.15/0 & 3.02/3.42/3.60/0 & 2.81/3.25/3.50/0 & 2.65/3.03/3.16/0 & 2.37/2.67/2.78/0 \\
 & 0 & 1920 & 5.22/5.76/6.10/0 & 5.00/5.34/5.51/0 & 4.81/5.19/5.45/0 & 5.03/5.42/5.62/0 & 4.67/4.99/5.07/0 \\
 & 1 & 800 & 3.77/5.24/5.62/0 & 3.50/4.79/5.13/0 & 3.31/4.67/5.02/0 & 3.02/3.72/3.86/0 & 2.71/3.38/3.50/0 \\
 & 1 & 1920 & 5.52/5.83/5.88/0 & 5.04/5.48/5.60/0 & 4.93/5.32/5.46/0 & 5.11/5.49/5.69/0 & 4.68/5.13/5.33/0 \\
 & 2 & 800 & 3.51/4.35/4.89/0 & 3.22/3.96/4.64/0 & 3.07/3.83/4.46/0 & 2.90/3.28/3.53/0 & 2.58/2.92/3.22/0 \\
 & 2 & 1920 & 5.25/5.53/5.62/0 & 4.90/5.20/5.36/0 & 4.77/5.02/5.19/0 & 5.05/5.30/5.48/0 & 4.64/4.91/5.10/0 \\
\midrule
DrJohnson & 0 & 800 & 2.60/9.26/10.97/9 & 2.80/8.45/9.89/6 & 2.34/8.23/9.89/5 & 2.43/5.42/5.65/0 & 2.22/4.72/4.87/0 \\
 & 0 & 1920 & 4.78/10.00/10.61/13 & 5.00/8.81/9.47/10 & 4.62/8.84/9.46/10 & 4.75/9.45/10.13/12 & 4.59/8.43/9.12/8 \\
 & 1 & 800 & 2.08/3.32/3.51/0 & 2.46/3.52/3.70/0 & 1.98/3.06/3.24/0 & 2.05/2.95/3.28/0 & 1.97/2.77/3.05/0 \\
 & 1 & 1920 & 4.48/5.49/6.93/0 & 4.69/5.60/7.04/0 & 4.33/5.33/6.71/0 & 4.43/5.47/6.90/0 & 4.32/5.29/6.69/0 \\
 & 2 & 800 & 3.28/6.53/8.07/0 & 3.42/6.04/6.97/0 & 2.99/5.70/6.95/0 & 2.99/5.32/5.74/0 & 2.73/4.57/4.96/0 \\
 & 2 & 1920 & 5.86/9.44/10.14/12 & 5.77/8.75/9.13/7 & 5.47/8.59/9.07/7 & 5.76/9.23/9.77/11 & 5.38/8.46/8.93/6 \\
\bottomrule
\end{tabular*}
\par\smallskip\raggedright Each cell gives P50/P95/P99 in milliseconds, followed by the number of missed frames out of 97. Clip uses 2~px. Each scene contributes three physical camera routes, each measured at both widths.
\end{table}
\endgroup

\subsection{Near-Plane Clipping Consistency}
\label{app:near-plane-clipping}
\label{app:arxiv-S4.4}

Across 1,040 trajectory conditions spanning 12 scenes, we compare gsplat and Graphdeco outputs on identical camera poses. Aligning gsplat's near clipping plane from 0.01 to 0.2 reduces the mean absolute RGB difference by more than $10^{-5}$ on at least one pose in 508 conditions, while zero conditions experience an error increase exceeding that threshold.

\end{document}